# Thermal conductivity of polymers and polymer nanocomposites

Congliang Huang,[1,2] Xin Qian,[2] and Ronggui Yang[2,3, *]

[1] School of Electrical and Power Engineering, China University of Mining and Technology, Xuzhou 221116, P. R. China.

[2] Department of Mechanical Engineering, University of Colorado, Boulder, Colorado 80309, USA.

[3] Materials Science and Engineering Program, University of Colorado, Boulder, Colorado 80309, USA.

*Email: Ronggui.Yang@Colorado.Edu

**Abstract:** Polymers are widely used in industry and in our daily life because of their diverse functionality, light weight, low cost and excellent chemical stability. However, on some applications such as heat exchangers and electronic packaging, the low thermal conductivity of polymers is one of the major technological barriers. Enhancing the thermal conductivity of polymers is important for these applications and has become a very active research topic over the past two decades. In this review article, we aim to: 1). systematically summarize the molecular level understanding on the thermal transport mechanisms in polymers in terms of polymer morphology, chain structure and inter-chain coupling; 2). highlight the rationales in the recent efforts in enhancing the thermal conductivity of nanostructured polymers and polymer nanocomposites. Finally, we outline the main advances, challenges and outlooks for highly thermal-conductive polymer and polymer nanocomposites.

# Table of Contents

**Nomenclature**

| | | | |
|---|---|---|---|
| 0D | 0- dimensional | PA6 | polyamide-6 |
| 1D | 1- dimensional | PBX | polymer-bonded explosives |
| 2D | 2- dimensional | PC | polycarbonate |
| 3D | three dimensional | PDA | polydopamine |
| $Al_2O_3$ | aluminum oxide | PE | polyethylene |
| AlN | aluminum nitride | PEEK | poly(ether-ether-ketone) |
| BB | Bottlebrush | PEG | Poly(ethylene glycol) |
| BN | boron nitride | P-GC | phenyl-aminated GO and CNT hybrid fillers |
| BNNS | boron-nitride nanosheets | | |
| CE | cyanate ester | PHB | poly(3-hydroxylbutyrate) |
| CF | carbon foam | PI | polyimide |
| CNT | carbon nanotubes | PMMA | poly(methyl methacrylate) |
| EG | expanded graphite | PP | Polypropylene |
| E-GC | ethyl-aminated graphene-oxide and CNT hybrid fillers | PPS | polyphenylene sulfide |
| | | PS | polystyrene |
| EMT | effective medium theory | PVA | Poly(vinyl alcohol) |
| GF | graphene foam | PVDF | poly(vinylidene fluoride) |
| GNP | graphene nanoplate | PVP | Polyvinylpyrrolidone |
| GNR | graphene nanoribbons | PVPh | poly(4-vinyl phenol) |
| GO | graphene oxide | R-GC | raw GO and CNT hybrid fillers |
| GS | graphene sheet | $Si_3N_4$ | silicon nitride |
| H-bond | Hydrogen bond | SiC | silicon carbide |
| h-BN | hexagonal boron nitride | SR | silicone rubber |
| HDPE | high density polyethylene | SWCNT | Single-wall CNT |
| MD | Molecular dynamics | UHMW-PE | ultrahigh molecular weight polyethylene |
| MgO | magnesium oxide | | |
| MWCNT | Multi-wall CNT | vdW | van der Waal |
| P3HT | poly(3-hexylthiophene-2,5-diyl) | ZnO | zinc oxide |
| | | P(VDF-TrFE) | poly(vinylidene fluoride-trifluoroethylene) |
| PAA | Poly(acrylic acid) | | |

# 1 Introduction

Polymers and polymer composites are used ubiquitously in a wide range of industrial applications ranging from structural materials to electronics and in our daily life from chopsticks to trash bins due to their diverse functionality, light weight, low cost, and excellent chemical stability. However, the low thermal conductivity of polymers limits in their applications in some fields. For example, the low thermal conductivity of polymers can be one of the major technological barriers for the polymer-based flexible electronics due to the limited heat spreading capability. [1-3] If a polymer can be engineered with high thermal conductivity, polymeric heat spreaders and heat exchanger can be manufactured with superb features including structural compactness, light weight, resistance against corrosions, and ease of processing and low-cost, which could in turn find many applications in electronics, water and energy industry. [4,5] Thus, enhancing the thermal conductivity of polymers and polymer composites are of great interests.

Over the past two decades, with a better understanding of the fundamental heat transfer process at the micro-, nano- and even molecular- scales, there have been significant efforts devoting to enhancing the thermal conductivity of polymers and polymer nanocomposites, which are expected to enable a broader range of applications. In this review, we aim to: 1). systematically summarize the understanding on the physical mechanisms that controls the thermal transport in polymers by relating those to polymer chain morphology and inter-chain coupling; 2). highlight the rationales in the recent efforts in enhancing the thermal conductivity of nanostructured polymers and polymer nanocomposites.

The thermal conductivity of bulk polymers is usually very low, on the order of 0.1 - 0.5 $W·m^{-1}·K^{-1}$, which is due to the complex morphology of polymer chains. [6] Fig. 1(a) shows a typical structure of a polymer, which consists of crystalline domains where polymer chains are aligned periodically, and amorphous domains where the polymer chains are randomly entangled. The thermal conductivity of a polymer depends greatly on its morphology. When amorphous domains are dominant, vibrational modes in the polymer tend to be localized, resulting in a low thermal conductivity. It is therefore natural to expect that thermal conductivity can be enhanced by improving the alignment of polymer chains. Indeed many efforts have been devoted to align polymer chains to enhance the thermal conductivity, by using mechanical stretching, nanoscale templating and electrospinning. A thermal conductivity as high as 104 $W·m^{-1}·K^{-1}$ has been achieved for polyethylene (PE) after stretching with a draw ratio of 400 for nanofibers with a diameter of 50-500 nm and lengths up to tens of millimeters [7]. The thermal conductivity was shown to be enhanced for more than 20 times in polythiophene nanofibers with a fiber diameter of about 50 nm, prepared using templated electropolymerization [8]. Inspired by these experimental efforts, molecular dynamics simulations have been conducted to understand how nanoscale structures affect the thermal conductivity. As shown in Fig. 1(b), in addition to polymer chain alignment, the thermal conductivity of a polymer also depends on the structure of chains including backbone bonds and side chains, and the inter-chain coupling.

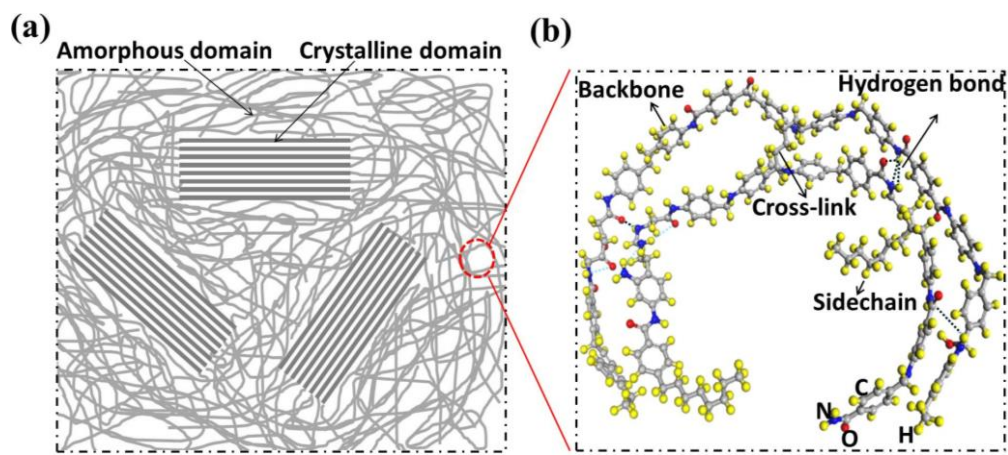

*Fig. 1 Schematic diagrams of a polymer: (a) the morphology of a polymer consisting of crystalline and amorphous domains; (b) structure of a polymer chain.*

In addition to engineering the morphology of polymer chains, another common method to enhance the thermal conductivity of polymers is to blend polymers with highly thermal conductive fillers. The progress of nanotechnology over the last two decades not only provides more diverse high thermal conductivity fillers of different material types and topological shapes but also advances the understanding at the nanoscale. Fig. 2 shows a sketch of a polymer nanocomposite to illustrate the thermal transport mechanisms. In general, there are two types of polymer nanocomposites depending on whether nano-fillers form a network or not. When the filler concentration is low, no inter-filler networks could be formed, as shown in Fig. 2(a). The thermal conductivity is essentially determined by the filler-matrix coupling, i.e, interfacial thermal resistance, and the concentration and the geometric shapes of fillers. When the filler concentration is large enough, high conductivity fillers might form thermally conductive networks, as shown in Fig. 2(b). Although nanocomposites with filler network could possess a higher thermal conductivity than that without a network, their thermal conductivity could still be low due to the large inter-filler thermal contact resistance. Recently, three-dimensional fillers, such as carbon and graphene foams, have drawn a lot of attention. The fundamental thermal transport mechanisms and recent synthesis efforts in both types of nanocomposites are reviewed.

This review article is organized as follows. In Section 2, we introduce the experimental progress on the enhancement of thermal conductivity by aligning polymer chains, and then review the methods to further tune the thermal conductivity by engineering chain structure and inter-chain coupling, as illustrated in Fig. 3(a). In Section 3, we discuss the thermal conductivity of polymer nanocomposites both with and without inter-filler networks, as illustrated in Fig. 3(b).

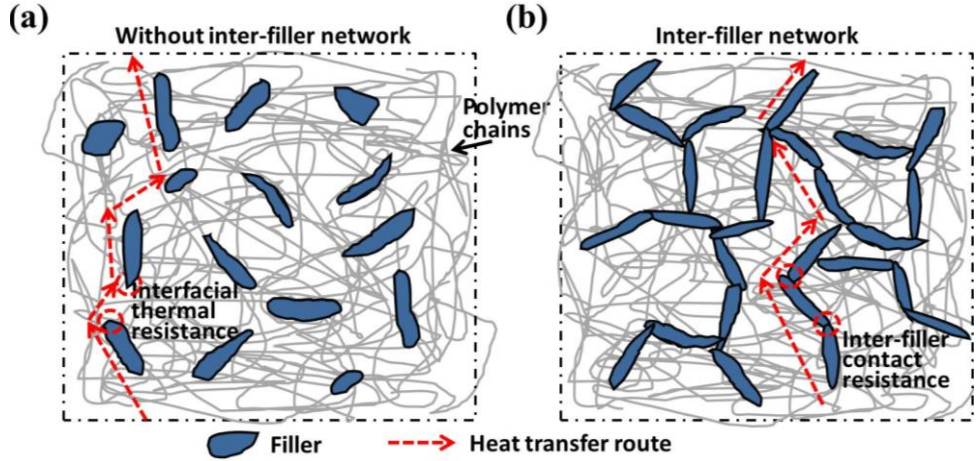

*Fig. 2 Schematic diagrams of polymer nanocomposites: (a) without inter-filler network; (b) with inter-filler networks. Thermally conductive pathway is identified with dash lines.*

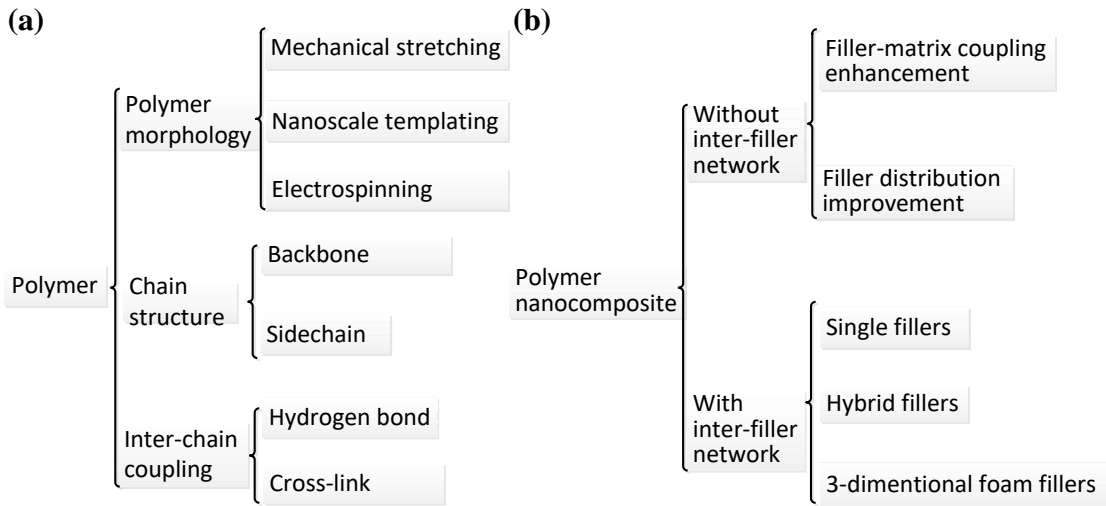

*Fig. 3 Physical mechanisms affecting the thermal conductivity of: (a) polymers; (b) polymer nanocomposites.*

## 2 Thermal conductivity of pristine polymers

In this Section, we discuss the dominant factors controlling the thermal transport in pristine polymers and the methods for manipulating the thermal conductivity through morphology control. One of the most intuitive methods for improving the thermal conductivity is to improve the order of chain alignment (chain morphology), which has been experimentally demonstrated. This method is discussed in Section 2.1. Apart from the chain morphology, the chain structure also plays an important role in determining the thermal conductivity. Methods for manipulating thermal transport by engineering the chain structure including backbone and side chains are discussed in Section 2.2. In Section 2.3, we further discuss the methods by enhancing inter-chain coupling through introducing strong interactions like hydrogen bonds (H-bonds) and cross-linkers formed by covalent bonds to tune the thermal conductivity of polymers, as compared to the weak van der Waals (vdW) in most of polymers.

## 2.1 Alignment of chain orientations

A typical semi-crystalline polymer contains crystalline domains of aligned chais, and the amorphous domains with randomly twisted and entangled chains. The lack of periodicity in the amorphous domains severely localizes the vibrational modes, thereby suppressing the thermal conductivity. Increasing the chain alignment is therefore expected to enhance the thermal conductivity of polymers. In crystalline polyethylene (PE) nanofibers, the measured thermal conductivity could be as high as 104 $W \cdot m^{-1} \cdot K^{-1}$ [7]. There have already been some experimental methods demonstrated to improve the degree of the chain alignment. For example, the chain alignment could be enhanced by thermal annealing [9-11]. Experimental studies on the poly(vinylidene fluoride-trifluoroethylene) [P(VDF-TrFE)] polymer showed that when the polymer film was annealed above its melting temperature (421 K), a 300-400 % increase of thermal conductivity occurs because of the chain alignment [12]. It is also unveiled that the geometric constraints at the substrate-P(VDFTrFE) interface have an important role on the formation of aligned chains in the process of melt-recrystallization [12], which has also confirmed by other works [13,14].

The mechanisms of how annealing and geometric constraints of the substrate affect chain alignment of polymers is still not well understood and thus it is difficult to further improve the chain alignment. In this Section, some experimental methods which have been successfully applied to enhance the chain alignment, such as the mechanical stretching, nanoscale templating and electrospinning, are summarized in Sections 2.1.1-2.1.3.

## 2.1.1 Mechanical stretching

Mechanical stretching could significantly increase the thermal conductivity of polymers due to the increased order of chain orientation. The first demonstration was performed by Choy et al. [15-17], who found that the thermal conductivity of the ultrahigh molecular weight polyethylene (UHMW-PE) could exceed 40 $W \cdot m^{-1} \cdot K^{-1}$ when the drawing ratio reaches beyond 300 [17], as shown in Fig. 4(a). Using a two-stage heating method, Shen et al. [7] obtained a higher drawing ratio (400) and produced a PE nanofiber with a diameter of 50-500 nm where a higher thermal conductivity of 104 $W \cdot m^{-1} \cdot K^{-1}$ was obtained, as illustrated in Fig. 4(a). The authors attributed the enhancement of the thermal conductivity to the improved fiber crystallinity. Some other experimental works also confirmed the enhancement thermal conductivity due to the enhanced chain orientation through mechanical stretching [18,19].

However, a higher crystallinity does not always lead to a higher thermal conductivity, because the thermal conductivity depends on the overall chain alignment in the polymer rather than the portion of the extremely aligned chains (crystalline). An example supporting this understanding is shown in Fig. 4(b). The thermal conductivity of a thermally stretched UHMW-PE microfiber approaches 51 W·m$^{-1}$·K$^{-1}$, while its crystallinity is reduced from 92% to 83% during the stretching [20]. Such increased thermal conductivity is due to the improved chain alignment in the amorphous domains.

To theoretically understand the enhanced thermal conductivity by mechanical stretching, there have been quite some research works using atomistic simulations. For example, using MD simulations, Liu and Yang [21] showed that the thermal conductivity of PE increases when the polymer is stretched slowly. Such thermal conductivity enhancement is found to be strongly correlated with the orientation order parameter [22] which describes the change of chain conformation. However, the tensile extension not always leads to an enhanced thermal conductivity. For example, Wang and Lin [23] found that the thermal transport in cumulene is relatively independent of the strains, as shown in Fig. 4(c). This can be explained by the two competing factors during stretching, one is the increased phonon lifetime due to the increased order of chain morphology and the other is decreased group velocity due to the strains.

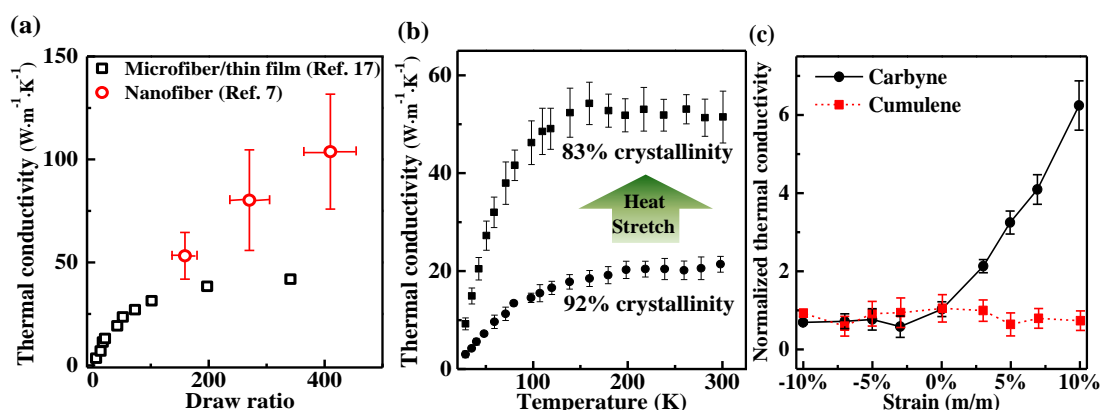

Fig. 4 Thermal conductivities under mechanical stretching: (a) effect of draw ratios in UHMW-PE [7]. Copyright 2010 Nature Publishing Group; (b) influence of crystalline in UHMW-PE. [20] Copyright 2017 American Chemical Society; (c) Tensile effects on thermal transport in carbine and cumulene [23]. copyright 2015 Nature Publishing Group.

### 2.1.2 Nanoscale templating

Nanoscale templating is another way to align polymer chains. Polymers are melted and infiltrated in a porous template such as porous anodic alumina. Removing the porous anodic alumina template using NaOH aqueous solution left with an array of polymer nanofibers. The alignment of the polymer chains is improved due to the flow of polymer melt in the nanoporous template, and the thermal conductivity can therefore be enhanced along the axial direction of a fiber.

In the work by Singh et al. [8], aligned arrays of polythiophene nanofibers were electro-polymerized vertically inside the nano-channels of anodic alumina templates, as shown in Fig. 5(a). The length of the polythiophene fibers is tuned by the charge passing through the electrochemical cell and the diameter is controlled by the diameter of the pores in the template. The thermal conductivity of polythiophene nanofibers is therefore improved to be 4.4 W·m$^{-1}$·K$^{-1}$, more than 20 times higher than that of the bulk polymer, because of improvement of the chain alignment as schematically shown in Fig. 5(b).

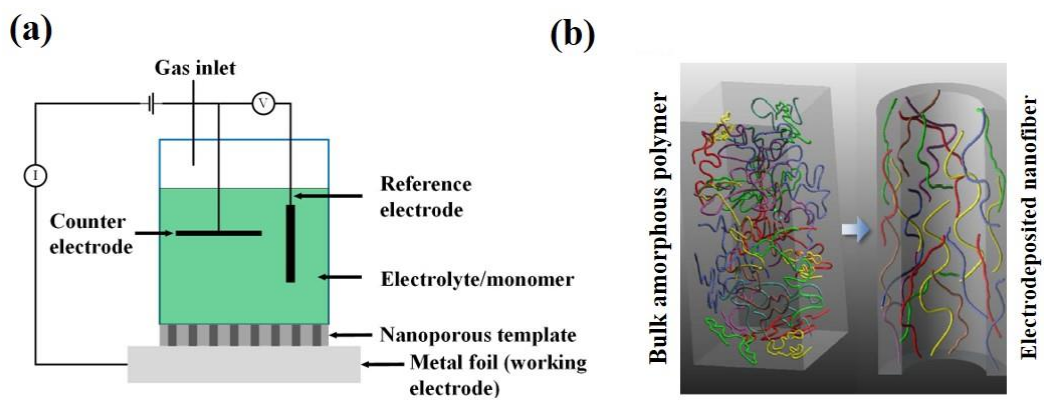

*Fig. 5 Nanoscale templating method: (a) schematic diagram of experimental setup for template guided electrochemical synthesis of nanofibers; (b) chain orientation comparison between bulk polymer (left) and electropolymerized nanofiber (right). [8] Copyright 2014 Nature Publishing Group.*

The thermal conductivity could be increased by decreasing the fiber diameters with a nanoscale templating method, because of the decreased entangle possibility of the chains. The thermal conductivity of approximately 7 W·m$^{-1}$·K$^{-1}$ was achieved for both the 200-nm-diameter high-density PE (HDPE) fibers and the 100-nm-diameter poly(3-hexylthiophene-2,5-diyl) (P3HT) fibers [24]. The thermal conductivity of the 200-nm-diameter HDPE nanofiber can be further improved to 26.5 W·m$^{-1}$·K$^{-1}$ using the nanoporous template wetting technique. The authors attribute this large increase of thermal conductivity to the integrative effects of shear rate, vibrational perturbation, translocation, nano-confinement and crystallization.

### 2.1.3 Electrospinning

Electrospinning is a nanofiber production technique which uses electrostatic forces to draw charged threads of polymeric solution. A polymer is dissolved in a solution and held in a reservoir. Usually a syringe with a sharp tip is used for dispensing the polymer. By applying an electrostatic force between the tip and a grounded collector plate, polymer solution is drawn to fibers with nano-scale diameters. [25] There are two experimental parameters affecting the chain morphology of electrospun polymers. The first parameter is the strength of the electric field during the electrospinning process. It is reported that strengthening the electric field is beneficial to enhance the chain alignment, and thus the thermal conductivity of the electrospun PE nanofibers [26,27]. According to the degree of polymer chain alignment in an electrospun fiber, Canetta et al. [28] summarized the possible orientations of the chains in a polymer as shown in Fig. 6. To enhance the thermal conductivity, it is preferable that the chains are mainly aligned along the axis of the fiber as shown in Fig. 6(a). The other parameter affecting the alignment of polymer chains is the jet speed. As the evaporation is faster on the outside of the jet in electrospinning, a core-shell morphology as shown in Fig. 6(b) could be formed in electrospun fibers [29], which is helpful for enhancing the thermal conductivity. Sometimes, the super-molecular morphology as shown in Fig. 6(c) could be also formed to enhance the elastic modulus of electrospun fibers [30], which is usually related to a higher thermal conductivity.

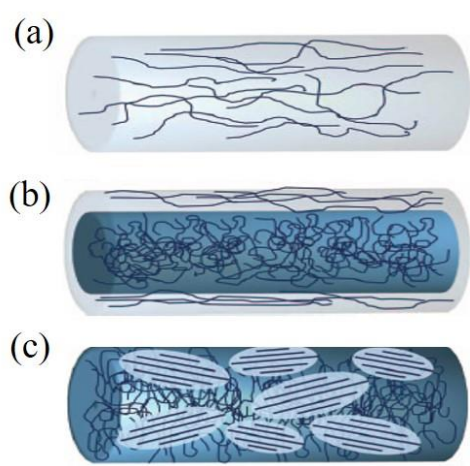

*Fig. 6 Orientations of polymer chains: (a) a nanofiber with preferable aligned chains; (b) a core-shell morphology with the shell formed by aligned chains and core comprised of randomly oriented chains; (c) a super-molecular morphology with aligned-chain grains filled in a randomly oriented chain matrix. [28] Copyright 2014 American Institute of Physics.*

Figure 7 shows the enhanced thermal conductivity of polymers with chain alignment through different processing techniques. The thermal conductivity of the electrospinning PS nanofibers by Canetta et al. [28] is found to be between 6.6 and 14.4 $W·m^{-1}·K^{-1}$, a significant increase above the typical thermal conductivity (~ 0.15 $W·m^{-1}·K^{-1}$) of bulk PS due to the preferential alignment of molecular chains, along with the reduction in defects and voids compared to bulk. Zhong et al. [31] reported enhanced thermal conductivity (1-2 $W·m^{-1}·K^{-1}$) of Nylon-11 nanofibers fabricated by electrospinning and post-stretching. They revealed that the crystalline morphology plays an important role to enhance the thermal conductivity in addition to the chain alignment. In general, Fig. 7 shows that the thermal conductivity increases with the decreasing nanofiber diameters, because the fiber surface could limit the random orientation of polymer chains [28].

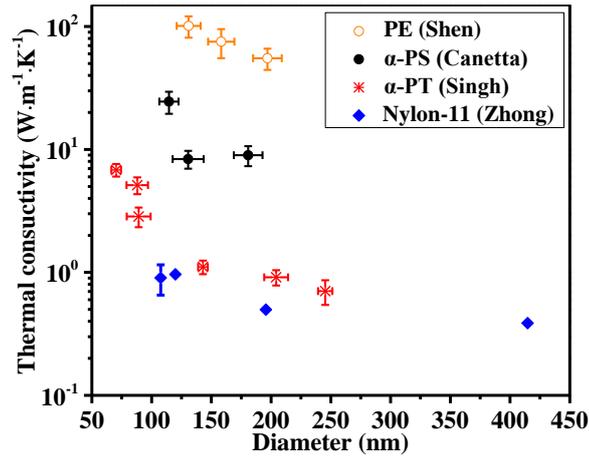

*Fig. 7 Measured thermal conductivity for nanofibers of PE [7], Nylon-11 [31], amorphous PS (a-PS) [28], and amorphous polythiophene (a-PT) [8]. [28] Copyright 2014 American Institute of Physics.*

## 2.2 Atomic structure of polymer chains

The structure of polymer chains could also play an important role in the thermal conductivity. The bond stretching and the angular bending strengths of the backbone influence the chain entanglement, while the strength of dihedral bending can also greatly affect the rotation of chain segments. Both the chain entanglement and the rotation of chain segments can in turn affect the thermal conductivity while the side-chain branching out of the backbone has a role as well.

### 2.2.1 Backbone

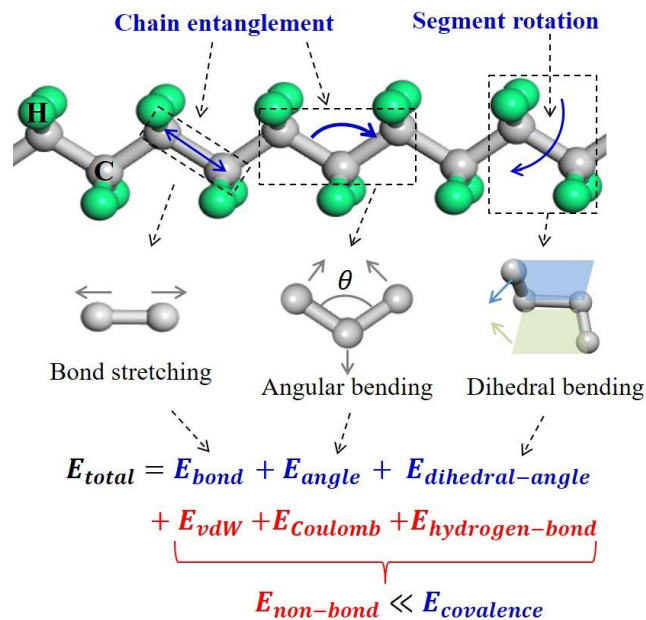

$$E_{total} = E_{bond} + E_{angle} + E_{dihedral-angle} + E_{vdW} + E_{Coulomb} + E_{hydrogen-bond}$$

$$E_{non-bond} \ll E_{covalence}$$

*Fig. 8 Effect of backbone structure of a polymer chain on the thermal conductivity, illustrated using a PE polymer chain as an example.*

In Fig. 8, we use a PE polymer chain as an example to illustrate the backbone structure and the conformation energy of the chain contributed by different vibrational modes. The total conformation energy of a polymer chain can be separated into the conformation energy of the covalent bonds and the non-bonding energy due to van der Waals (vdW) interactions, Coulomb interactions and hydrogen bonds. The non-bonding force is usually much smaller than the covalent force and thus is not as important as covalent bonds in affecting the thermal conductivity of a continuous single chain, we therefore focus our discussions on the conformation energy associated with covalent bonds. As shown in Fig. 8, the covalent bonding energy can be decomposed into three parts according to different vibrational modes. The first part is the two-body stretching energy $E_{bond}$, due to the stretching of a bond connecting two atoms. In addition to the two-body bond stretching, many body interactions also contribute to the conformation energy, and usually up to four body interactions are considered, which are the three-body angular bending motion, $E_{angle}$, and the four-body dihedral bending motion, $E_{dihedral-angle}$. The angular bending involves three atoms connected by two neighboring covalent bonds, and the three atoms have a scissoring like motion that changes the angle between the two covalent bonds. A dihedral bending motion involves four atoms connected by three nose-to-tail covalent bonds. The dihedral angle is the angle between the two planes sharing a covalent bond in the middle and the atom at two ends of the atomic quadruplet. The $E_{dihedral-angle}$ describes the energy barrier to overcome if a polymer segment is rotated. A larger bond-stretching energy $E_{bond}$ or/and angular bending energy $E_{angle}$ means a stiffer backbone, and a larger dihedral-angle energy $E_{dihedral-angle}$ tends to suppress the segmental rotation. In the following, the effect of the backbone stiffness (bond-stretching and angle-bending strengths) and the dihedral-angle-torsion stiffness on thermal transport in polymer chains are discussed.

Generally, higher stiffness of bond stretching and angular bending leads to a higher thermal conductivity. One well-known example is PE, which has very high intrinsic thermal conductivity due to the strong bonding, where both the bond stretching and bond bending are very stiff [32,33]. Fig. 9 shows a more detailed thermal conductivity dependence on the stiffness of the polymer chains [34]. Generally, the existence of double -C=C- bond with sp$^2$ hybridization results in higher thermal conductivity due to two reasons: (1) The bonding energy of –C=C- bond is 2.8 times that of a -C-C- with sp$^3$ hybridization; (2) Consecutive sp$^2$ bond forms delocalized conjugated π-bond, which constrains the atoms in the conjugated π-bond to be in the same atomic plane. As a result, the conjugated π-bond greatly increases the stiffness of segmental rotation. One example is shown in Fig. 9(a), where the thermal conductivity of π-conjugated polyacetylene is higher than that of PE. [35] A recent work by Xu *et al.* [36] found that the existence of conjugated -C=C- in poly(3-hexylthiophene) (P3HT) can simultaneously achieve efficient phonon transport along the chains and strong noncovalent inter-chain interaction due to the π-π stacking. As a result, thermal conductivity as large as 2.2 W·m$^{-1}$·K$^{-1}$ is measured in oxidative chemical vapor deposition (oCVD) grown thin films. This work shows that polymers with conjugated bonds are promising candidates for achieving higher thermal conductivity. Conjugated π-bond is also found in aromatic rings as shown in Fig. 9(b), where the thermal conductivity of chains with aromatic-backbone structures is much higher than that with aliphatic-backbone structures. In addition, the type of carbon-carbon bonding, the thermal conductivity of polymer chains can also be tuned by replacing the hydrocarbon functional groups in the backbone with other atomic species. Fig. 9(c) shows the effect of replacing –CH$_2$- groups in the backbone with O atoms. Because the -CH$_2$-O- bonding energy (~335 kJ/mol) is lower than –CH$_2$-CH$_2$- (~ 350 kJ/mol) and O atom is heavier than –CH$_2$- group, the thermal conductivity of poly(ethylene oxide) is much lower than that of both PE and poly(methylene oxide).

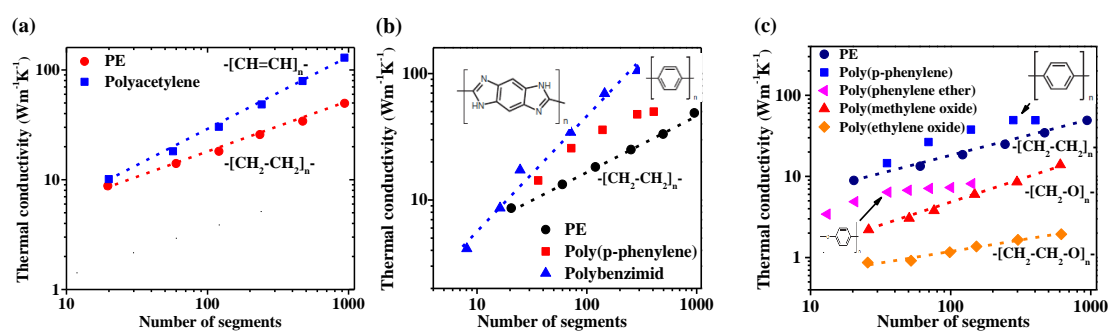

Fig. 9 The thermal conductivity dependence of bond strength: (a) effect of double bonds; (b) effect of aromatic backbone; (c) effect of bond-strength disorder. [34] Copyright 2012 American Physical Society.

The effect of the dihedral-angle stiffness on the thermal conductivity has also been studied using MD simulations performed in a model system of bulk PE amorphous polymer [37]. The energy constant ($K_1$) of the dihedral angle energy which stands for the dihedral-angle stiffness is systematically changed, and the resultant structures and thermal conductivity are shown in Fig. 10. When $K_1$ is increased, the radius of gyration increases, which means more extended chains (Fig. 10a), and the persistence length also increases, which suggests more straight chains (Fig. 10a). Because of more extended and straight chains, the thermal conductivity in all three orthogonal directions (X, Y and Z) increase with increasing $K_1$, as shown in Fig. 10(b). Further thermal conductivity decomposition analysis reveals that thermal transport through covalent bonds dominates the thermal conductivity over other contributions from the non-bonding vdW interactions and the translation of molecules (Fig. 10c). The contribution of the non-bonded vdW force to the thermal conductivity is also enhanced by the increase of energy constant ($K_1$) of the dihedral angle energy, because of the shortened inter-atomic distance.

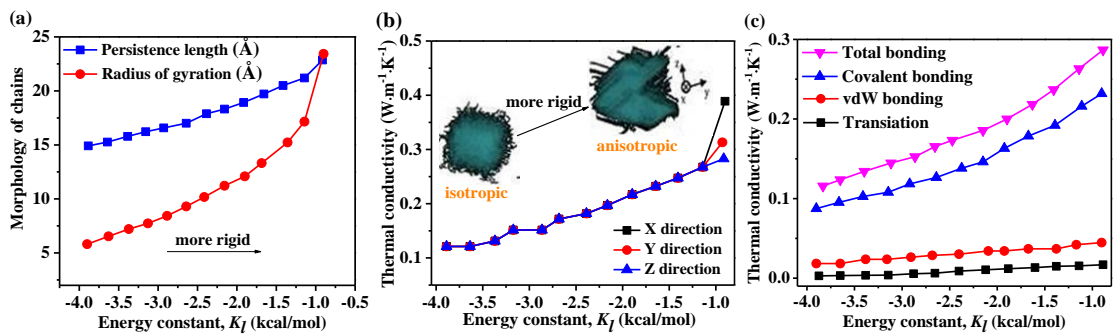

*Fig. 10 Effects of the dihedral-angle stiffness: (a) persistence length and radius of gyration of chains; (b) thermal conductivity; (c) contributions of covalent bonding and non-covalent bonding on the thermal conductivities along X direction. [37] Copyright 2016 American Chemical Society.*

**2.2.2 Side-chains**

Side-chains are functional groups branching out of the backbones [38]. Since side-chains change the topology of the polymer chains, the vibrational dynamics and the thermal conductivity are both expected to change when different types of side chains are introduced. Here we discuss the effect of side-chains on the thermal conductivity of polymers.

Thermal conductivity of a polymer chain is found to decrease when side-chains are introduced, and usually a heavier side-chain leads to a lower thermal conductivity one [38], as shown in Fig. 11. It also shows that the thermal conductivity is decreased when the distance between two neighboring side-chains is decreased from 100 segments, to 75 and then to 50 segments. The decreasing trend of the thermal conductivity with the number-density increase of the side-chains is further shown in Fig. 12(a). A larger number of side-chains could lead to a lower thermal conductivity for both aligned and fork arrangements of side-chains [38]. With the increase of the number density of side-chains, the thermal conductivity of a PE-ethyl chain finally converges to be only about 40 % that of the pristine PE chain [38].

A larger length of side-chains could also lead to a smaller thermal conductivity. Ma and Tian [39] reported that the thermal conductivity decreases with the increase of the side-chain length, which is attributed to the increased phonon scattering. The thermal conductivity become insensitive to the side-chain length when this length is larger than 10 segments, as that shown in Fig. 12(b). The mechanism of the reduced thermal conductivity caused by the side-chains could be understood as a result of phonon localization and phonon scatterings [38-39]. In bulk polymers, side chains are also found to decrease the thermal conductivity, because the crystallinity are found to decrease dramatically when side chains are introduced [39].

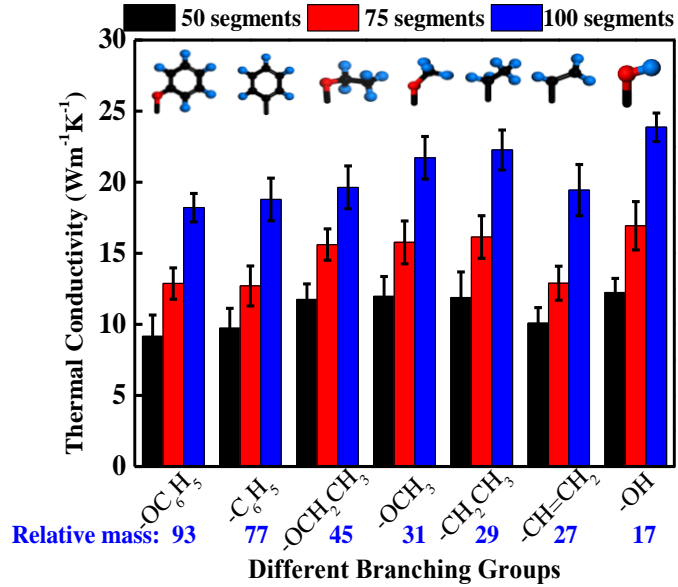

*Fig. 11 Thermal conductivities of PE chains branched with different side-chains. The side-chains are attached to backbone for every 50, 75, and 100 segments. The black, red and blue columns stand for different segments of PE backbones. [38] Copyright 2018 American Society of Mechanical Engineers.*

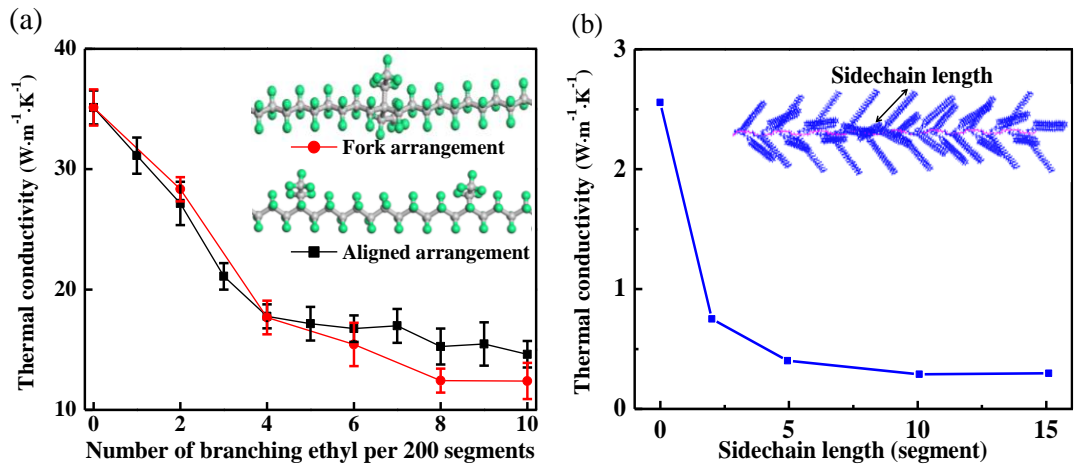

*Fig. 12 Effects of number density and length of side-chains on the thermal conductivity: (a) a single PE chain with different number density of side-chains [38]. copyright 2017 American Society of Mechanical Engineers; (b) a single bottlebrush chain with different side-chain length. Insets show the schematic of polymers [39]. Copyright 2017 American Institute of Physics.*

Interestingly, side-chains could also lead to an increase in thermal conductivity of polymers. Using the ultrafast laser based pump-probe technique, Guo et al. [40] showed that the measured thermal conductivity of conjugated polymers with linear and long side-chains increase by 160% compared to that with short side-chains. MD simulations further revealed that these linear and long side-chains tend to increase the structural order of the conjugated polymers, which results in an increased thermal conductivity. Chen et al. [41,42] also reported that the thermal conductivity of polyaniline can be enhanced by introducing functional groups due to the improved chain alignment. Side-chains can change the thermal conductivity in both ways which calls for more theoretical works to further understand the effect of side-chains on the thermal conductivity of polymers.

## 2.3 Inter-chain coupling

Weak inter-chain coupling has been considered as a bottleneck for thermal conduction in polymers. Enhancing inter-chain coupling strength could potentially lead to a higher thermal conductivity of polymers. For example, the thermal conductivity of polymer salts (Poly(vinylsulfonic acid Ca salt)) could be as high as 0.67 W·m$^{-1}$·K$^{-1}$ due to the relative strong electrostatic forces between the ions in different polymer chains, in comparison with the common 0.2-0.3 W·m$^{-1}$·K$^{-1}$ for most polymers[43]. In addition to the ionic inter-chain coupling, hydrogen bond (H-bond) is another type of relatively strong inter-chain coupling, which is 10-100 times stronger than the vdW interaction. Therefore, large concentration of H-bonds could also lead to a higher thermal conductivity. Cross-linking to form chemical bonds is apparently one of the most effective way in enhancing the thermal conductivity. Here we discuss the effect of H-bonds on the thermal conductivity in Section 2.3.1, and then the effect of covalent cross-links in Section 2.3.2.

### 2.3.1 Hydrogen bond

Hydrogen bond is a strong electrostatic interaction between the proton (or hydrated proton) and the lone electron pair(s) in O, N and F atoms. The H-bonds are beneficial to increase the thermal conductivity, because they could enhance the inter-chain coupling to form continuous thermal networks to provide more heat-transfer pathways. Therefore, a large density of H-bonds is usually desired for synthesizing high thermal conductive polymers [44,45]. Polymer blending is an effective way to increase the density of H-bonds because one component polymer can supply donating group while the other to supply accepting groups. Biopolymers in which the H-bond molecular group exist widely are usually applied in polymer blends, such as gelatin which possesses abundant carboxylic, amide, and amine groups, [46-48] lignin which owns abundant hydroxyl and aldehyde groups [49,50].

The H-bond density and the mixture ratio of polymer blends could strongly affect the thermal conductivity. To enhance the thermal conductivity, polymer blends not only need to be mixed uniformly to ensure a homogeneous distribution of H-bonds but also to allow polymers to intertwine within the radius of gyration to supply a continuous thermal network. This is systematically studied in Ref. [51], by mixing (poly(N-acryloyl piperidine) PAP with the H-bond donating polymer (poly(acrylic acid) (PAA), poly(vinyl alcohol) (PVA), or poly(4-vinyl phenol) (PVPh)). As shown in Fig. 13, there is an optimum composition of PAP at $\phi_{PAP} =$ 0.3 to enhance the thermal conductivity of polymer blends in Fig. 13(c) and (d), where $\phi_{PAP}$ is the mixture ratio of PAP. This could be explained by the fact that the backbone is mostly-extended at $\phi_{PAP} = 0.3$ while the H-bond density is above the percolation threshold to form a continuous thermal network (as illustrated in Fig. 13a).

How the H-bonds connect to the backbone has an important effect on the thermal conductivity. Fig. 13(c) and (d) show that the thermal conductivity of the polymer blends are higher than that of either component polymer, which can be explained by the strong H-bonds connected closely to the polymer backbones through the low-mass and short chemical linkers as illustrated in Fig. 13(b). On the contrary, the thermal conductivity of the PAP-PVPh polymer blends is lower than that of either constituent polymers in Fig. 13(e), which is because the H-bonds not directly connected with backbones as that illustrated in Fig. 13(b).

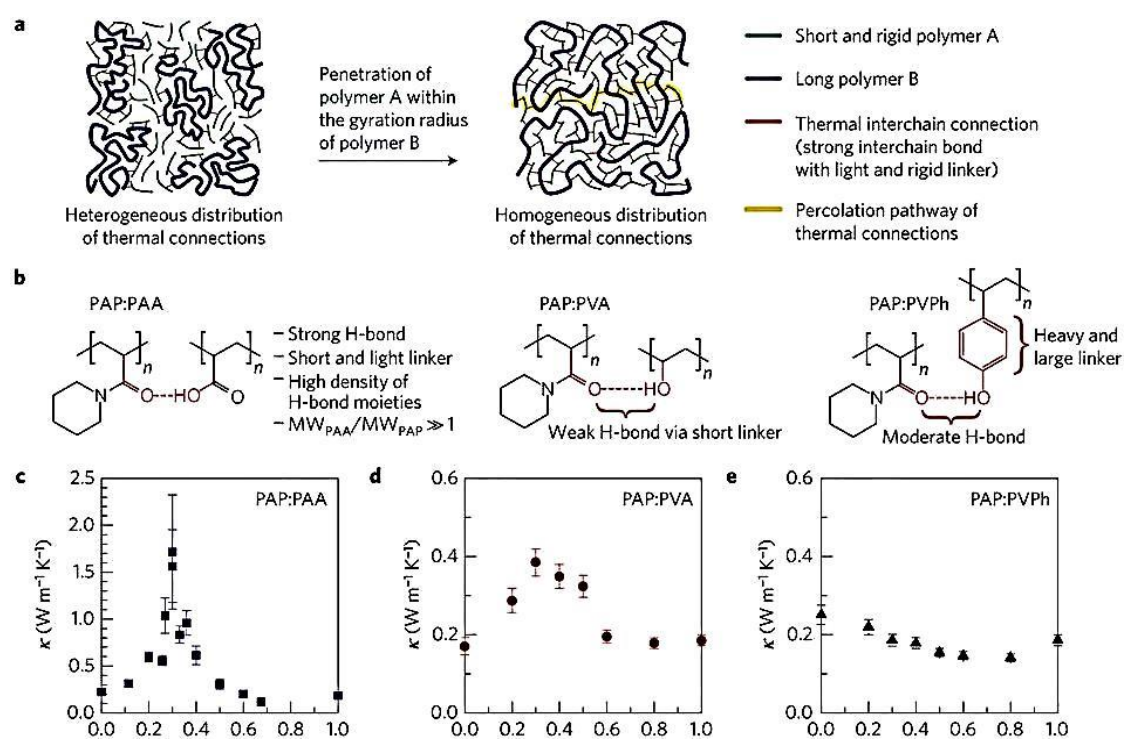

*Fig. 13 Thermal conductivity of polymer blends by engineering H-bond interactions: (a) Illustrations of heterogeneous (left) and homogeneous (right) distributions of H-bonds; (b) Inter-chain H-bonds (dashed lines); (c)-(e), measured thermal conductivities of PAP:PAA, PAP:PVA and PAP:PVPh polymer blends as a function of the mixture ratio of PAP ($\phi_{PAP}$). [51] Copyright 2015 Nature Publishing Group.*

In one of the most recent works, the importance in the formation of a continuous thermal network is further confirmed. The enhanced thermal conductivity of polymer blends is attributed to the H-bond-induced enlargement of chain coils and the continuous microstructures in polymers [52]. As illustrated in Fig. 14, by tailoring the distribution and the density of H-bonds among polymer blends with PVA and biopolymers (lignin, gelatin), an optimum concentration ratio of polymer blends to enhance the thermal conductivity is also observed. The highest thermal conductivity of PVA/lignin and PVA/gelatin blends are 0.53 W·m$^{-1}$·K$^{-1}$ at 2 wt% lignin loading and 0.57 W·m$^{-1}$·K$^{-1}$ at 5 wt% gelatin loading, respectively. Specifically, the thermal conductivity of the PVA polymer blended with 10 wt% lignin and 10 wt% gelatin reaches 0.71 W·m$^{-1}$·K$^{-1}$, which is much higher than that of the PVA/lignin and the PVA/gelatin polymer blends. This high thermal conductivity is attributed to the H-bond networks formed by these three polymer blends (PVA, lignin and gelatin) to supply more heat transfer pathways, as illustrated in Fig. 14.

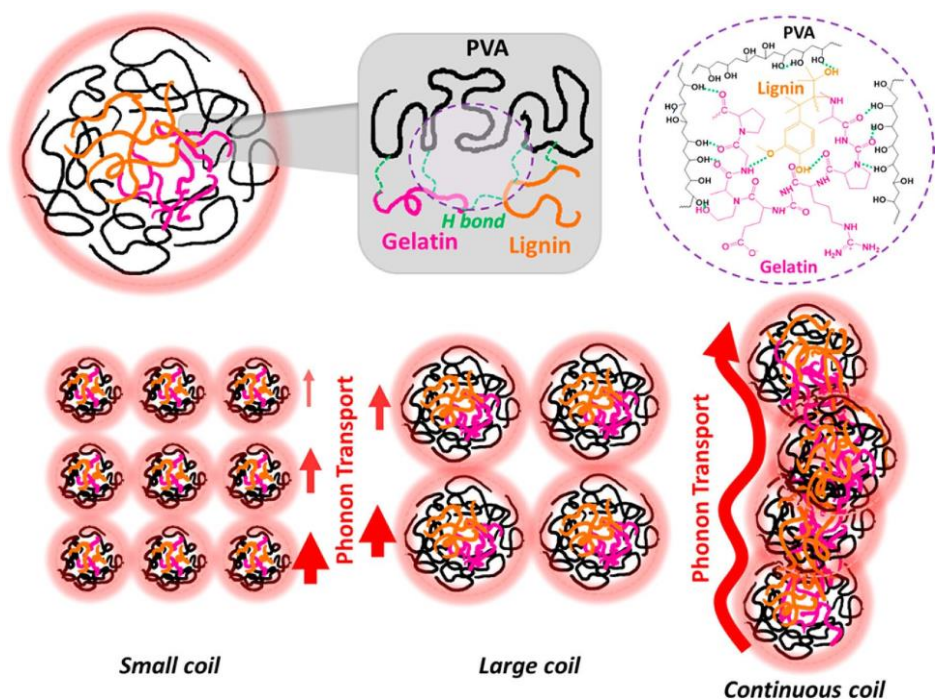

*Fig. 14 H-bonds in polymer blends and phonon transport for small, large, and continuous coil structures. [52] Copyright 2017 American Chemical Society.*

In addition to bridging different polymer chains to form heat-transfer networks (bridging effect), H-bonds could also serve as "soft grips" to suppress the segmental rotation of polymers, which reduces phonon scattering and thus leading to a higher thermal conductivity [53,54]. Through MD simulations of amorphous polymer blends, Wei et al. [55] proposed that locally ordered lamellar structures could form in the polymer blends due to the large inter-chain interactions and thus enhancing the thermal conductivity.

It is not always an effective way to tailor the thermal conductivity by modulating the density and the distribution of H-bonds. Experimental works show that the thermal conductivities of most polymer blends are still too low, 0.12 - 0.38 W·m$^{-1}$·K$^{-1}$, and the effect of H-bonds on the thermal conductivity is negligible [56]. Numerous polymers with strong hydrogen bonding still possess a quite low thermal conductivity, for example, 0.25 W·m$^{-1}$·K$^{-1}$ for nylon-6,6 [57]. The MD simulations by Zhang et al. [58] confirms that polymers with stronger inter-chain interactions (H-bonds, Columbic interactions, etc.) do not necessarily have higher thermal conductivity, because of possible weaker backbone bonds caused by inter-chain interactions. It was also revealed that H-bonds may increase the inter-chain phonon scatterings and thus suppressing the contributions of acoustic phonon modes to the thermal conductivity [59]. Similar effect of vdW interactions on the thermal conductivity was also observed in Ref. [60], where the thermal conductivity of PE strands decreases weakly as the chain number increases, because vdW interactions between chains introduce slightly more phonon scattering. Huang et al. [61] even attributed an exceptionally high thermal conductivity (up to 416 $W \cdot m^{-1} \cdot K^{-1}$, comparable to copper) of the dragline silk of spider to the break of the inter-chain H-bonds, which is needed for restructuring polymer chains.

**2.3.2 Crosslink**

Crosslinks can form efficient heat conduction pathways and networks by connecting polymer chains with strong covalent bonds. It is natural to expect that the thermal conductivity would increase with the increasing number of crosslinks in the polymer network [62,63]. For example, Tonpheng et al. [64] experimentally achieved a 50 % enhancement of the thermal conductivity for the PE polymer at a higher cross-linking densities. Using MD simulations, Kikugawa et al. [44] reported that the crosslink could significantly increase the thermal conductivity of the PE polymer and the increase of the thermal conductivity is proportional to the crosslink density.

It might be intuitive to think that the thermal conductivity enhancement is mainly due to the strong covalent bonds. However, recent MD simulations by Rashidi et al. [65] showed that the enhanced thermal conductivity with crosslinks cannot be explained completely by only considering covalent bonds. As shown in Fig. 15, the covalent bonds contribute to only 20% of the total thermal conductance for all number density of cross-links. More interestingly, the nonbonding interactions not only contribute to the majority of total conductance, but also their contribution of the total conductance increases with the increasing density of crosslinks. In addition to the covalent bonds connecting different chains, another effect of the crosslinks is to bring the polymer chains closer to each other. As a result, the non-bonding coupling (vdw, Coulombic or H-bonds) becomes stronger when there are more crosslinks in the polymer, which in turn significantly enhances the thermal conductivity.

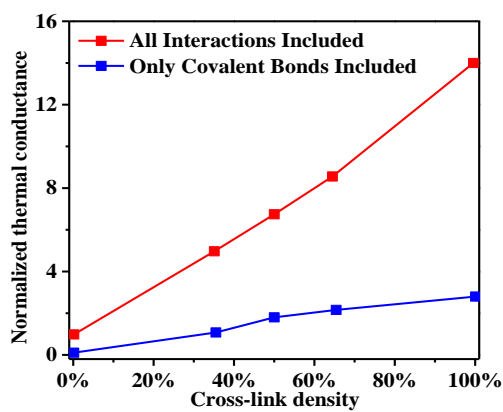

*Fig. 15 Thermal conductance in two PMMA chains cross-linked by $CH_2$, normalized by that with a cross-link density of 0, where the thermal conductance is defined as the heat flow divided by temperature difference. [65] Copyright 2017 American Chemical Society.*

Interestingly, crosslinks could also counter-intuitively decrease the thermal conductivity of polymers. For example, Yu et al. [66] experimentally showed a 30 % reduction in the thermal conductivity of HDPE, because the cross-links in HDPE suppressed the crystallization of polymer molecules. Although the crosslinks could enhance the inter-chain coupling strength, these crosslinks could also break the periodicity along the polymer chains, which could result in strong phonon scattering. For example, MD simulations performed by Ni et al. [67] suggest that a 10 % crosslinking in PE polymer could result in a 44.2 % reduction in the thermal conductivity along the chains. How to engineer the cross-link to tune the thermal conductivity of a polymer is still an open question.

## 3  Thermal conductivity of polymer nanocomposites

In addition to engineering polymer chains and morphology at the atomic/molecular level, the thermal conductivity of polymers can also be enhanced by adding highly thermal conductive fillers. Although the composites incorporated with macro- or micro-fillers have already been widely studied for close to a century using the effective medium theory, [68-71] the thermal conductivity of composites filled with nano-fillers can be quite different and not well understood yet. Different from the macro- or micro-scale composites, the large specific surface area of nano-fillers can lead to large contribution of interfacial thermal resistance in a nanocomposite. In addition, it is very challenging to mathematically describe the heat conduction if nano-fillers form a network in a nanocomposite. The thermal conductivity is determined not only by the polymer matrix and the fillers, but also the interaction between filler and matrix and among fillers (the filler network). In the following, we firstly introduce the influence of fillers on the thermal conductivity, and then discuss the thermal transport mechanism of different filler networks.

The commonly used fillers in nanocomposites can be categorized into the metallic, ceramic and carbonous fillers. The metallic fillers, such as copper nanoparticles [72,73], copper nanowires [74,75], aluminum fibers [76], silver particles [77], gold and palladium powders [78,79] could be used to enhanced the thermal conductivity of a polymer nanocomposite. However, these metallic fillers may also lead to an increase of electrical conductivity, which prevents their applications with electrical insulation requirement. Although the electrical conductivity of metallic fillers could be somewhat tailored by oxidation or surface treatment [80,81], the high thermally conductive ceramic fillers is more preferable for not only their electrical insulation property but also thermal stabilities. Typical high thermal conductivity ceramic nano-fillers are magnesium oxide (MgO) [82-84], aluminum oxide ($Al_2O_3$) [85-88], silicon nitride ($Si_3N_4$) [89-91], silicon carbide (SiC) [92-94], zinc oxide (ZnO) [95], aluminum nitride (AlN) [96-99], and boron nitride (BN) [100-104]. Compared with the metallic and ceramic fillers, nanostructured carbon fillers have attracted more intensive interests because of their high thermal conductivity. For example, expanded graphite (EG) has a thermal conductivity of about 300 $W·m^{-1}·K^{-1}$ [105], graphene nanoplate (GNP) possesses a thermal conductivity as high as 1000-5000 $W·m^{-1}·K^{-1}$ [106-110], and CNTs are regarded as the most promising candidates owing to their high mechanical strength, chemical stability and high thermal conductivity of 1000~3000 $W·m^{-1}·K^{-1}$ [111-114].

However, the known high thermal conductivity of individual CNT and GNP does not easily translate to high thermal conductivity nanocomposites. Until now, the thermal conductivity of CNT-polymer nanocomposite is still much lower than that with the commonly used metallic or ceramic fillers like aluminum or silicon carbide. There could be two possible reasons: 1). The quality of individual CNT and GNP used in nanocomposites can be quite different from those prepared for individual characterization of thermal properties. CNTs usually possess some defects which could greatly reduce its effective thermal conductivity, [115] such as Stone-Wales defect [116], doping or vacancy defects [117], and inter-tube junction [118,119]. Besides the intrinsic defects of CNT, the kinks, twists and waviness formed in the CNTs could also reduce the aspect ratio of CNT, and thus lead to a lower thermal conductivity of CNT-in-epoxy composites than expected [120,121]. Similarly, the reduced thermal conductivity due to the waviness is also observed in GNP [122]. To remove such defects in CNTs and GNPs, surface modification methods have been developed to alter the morphology and defect density, such as acid [123] and plasma treatment [124-126]. 2). Interface thermal resistance across the nano-filler and polymer can greatly reduce the benefits from the high thermal conductivity nano-fillers. When the filler concentration is low and a filler network is not formed, the large interfacial thermal resistance between fillers and matrix usually results in a low thermal conductivity. Furthermore, the polymer molecules wrapping around the CNT or GNP could induce strain and shape change and thus reducing the thermal conductivity of the CNT and GNP [127].

Increasing the amount of fillers to form a network can significantly enhance the thermal conductivity [128-135]. However, a high concentration of fillers could compromise other important properties of a polymer, such as mechanical, electrical, and optical properties. Constructing a 3-dimensional (3D) filler network by uniformly distributing fillers is a promising method to improve the thermal conductivity with a relatively low concentration of fillers. However, there yet exists a clear theoretical understanding of heat transfer mechanism in nanocomposites with 3D filler networks. Most of the theoretical models are based on the effective medium theory (EMT) [136-142] or the percolation-based theory [143-145], which might not be useful for nanocomposites with 3D filler network. Many of the numerical methods, such as Monte Carlo method [146-151], finite element method [152,153] and lattice Boltzmann method [154,155], recently developed by the nanoscale heat transfer community might shed some lights in the physical understanding, while challenging to be used for the design. In the following, we discuss the polymer nanocomposites both without inter-filler network and with inter-filler network to highlight the current understanding on the thermal conductivity of polymer nanocomposites.

## 3.1 Nanocomposites without an inter-filler network

When there is no inter-filler network formed in a polymer nanocomposite, most of the thermal resistance comes from the polymer matrix and the interfacial thermal resistance between matrix and fillers. In this case, the thermal conductivity increases gradually with the increasing concentration of fillers, and usually does not exhibit a percolation behavior. [156,157] The concentration-dependent thermal conductivity of a nanocomposite without inter-filler networks could be separated into three regimes. In the regime of low filler concentration, the fillers are independent from each other, thus the thermal conductivity increases with the increasing concentration of fillers (first rise). However, with the further increase of filler concentration, the fillers tend to aggregate because of insufficient volume for a high concentration [158]. When the fillers aggregate, the contact area between the filler and the polymer matrix decreases dramatically, and the thermal conductivity also sharply decreases. Besides the interfacial thermal resistance, the changing morphologies of conducting fillers (i.e. long aspect ratio fillers such as nanowires and CNT can be bended) also exert negative effect on the thermal conductivity of fillers, thus reducing the effective thermal conductivity of the nanocomposites [170,159]. If the filler concentration is further increased, a second rise of the thermal conductivity could occur because the clusters formed by aggregation of fillers could contact with others to form heat transport pathway.

The thermal conductivity in the 'first rise' regime has been well studied and reasonably well understood, with the EMT, under the dilute limit, by taking into account of interfacial thermal resistance [68,160,161]. At a low filling concentration, the heat flux lines in a nanocomposite generated by one particle are not distorted by the presence of the neighboring particles because the distance between neighboring particles is much larger than their size. However, for higher particle concentrations the distance between neighboring particles can be of the order of the particle size or smaller and the interaction among particles have to be considered, which results in distortion in heat flux that is different from the prediction of the single particle assumption. To take into account the particle interactions, based on the differential effective medium theory [162], Ordonez-Miranda et al. [163-165] have developed a crowding factor model to predict the thermal conductivity of composites, where the crowding factor is determined by the effective volume fraction of fillers. Their model generalize other EMT models which are suitable for nanocomposites, but also give a good thermal conductivity prediction of nanocomposites with a high filling ratio of fillers.

As an example, the experimental thermal conductivities of MWCNT@PP are shown in Fig. 16 [166]. The thermal conductivity exhibits a non-monotonic trend as a function of the filler concentration. The first rise of thermal conductivity is attributed to a uniform filler distribution, while the aggregation dominates the following decrease of the thermal conductivity with the increase of volumetric loading of MWCNT. With 3 vol% MWCNT loading in the matrix, the thermal conductivity decreases because some MWCNTs have to bend or agglomerate to fit into the small packing volume. Similar decrease trend of thermal conductivity is also observed in Refs. [167-170], which is attributed to the increased interfacial thermal resistance caused by aggregation. The second rise of the thermal conductivity may result from the heat transport pathway formed among clusters which are formed by aggregation of fillers.

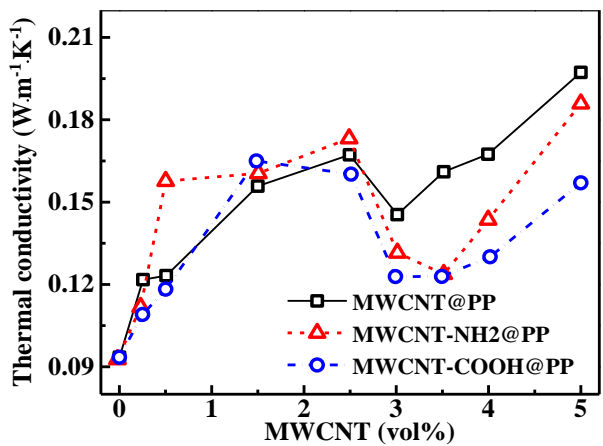

*Fig. 16 Experimental thermal conductivities of MWCNT@PP composites as a function of MWCNT volume fraction. [166] Copyright 2016 Elsevier Ltd.*

From the first rise to the decrease in the effective thermal conductivity, there exists a critical filler-matrix interfacial thermal resistance determining the turning point of the thermal conductivity. For example, the computational study of SWCNT@polymer shows that this critical SWCNT-matrix interfacial thermal resistance ($R_c$) is about $2 \times 10^{-7}$ m$^2$KW$^{-1}$, [171] as illustrated in Fig. 17. If the interfacial thermal resistance is smaller than $R_c$, the effective thermal conductivity of the nanocomposite would always increase with the volumetric ratio due to the large thermal conductivity of the SWCNTs. When the interfacial thermal resistance is larger than $R_c$, the thermal conductivity decreases with the increasing concentration of fillers. Unfortunately, the existing acoustic mismatch model and the diffuse mismatch model [172] could not accurately estimate the interfacial thermal resistance between the filler and the polymer matrix. MD simulations have been applied to elucidate the interfacial thermal resistance [173,174], where only the bonding or adhesion strength across the interface is considered. In a real nanocomposites, the interface thermal resistance between nano-filler and polymer can be dependent on many other factors including voids and molecule line-up at the interfaces.

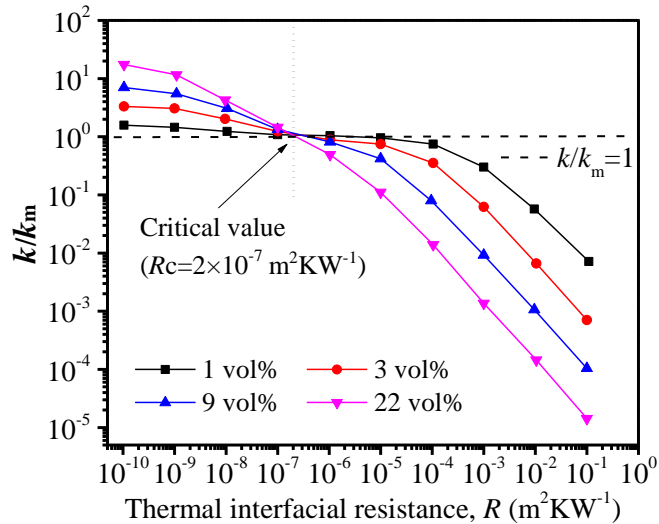

*Fig. 17 Thermal conductivity enhancement as a function of SWCNT-matrix interfacial thermal resistance ($R_c$) for SWCNT/polymer composites at SWCNT loadings of 1 vol%, 3 vol%, 9 vol% and 22 vol%. $k_m$ denotes the bulk thermal conductivity of the polymer matrix. [171] Copyright 2012 Elsevier Ltd.*

The addition of nano-filler can also affect the crystallization of the polymer matrix. Both SWCNT and MWCNTs can act as nucleation agents in increasing the polymer crystallization rate, which results in a larger thermal conductivity of polymer matrix and a reduced interfacial thermal resistance [175-177]. The crystallization behavior and the thermal conductivity of polymer matrix could be also enhanced by other kind of fillers, such as nanoparticles [178,179]. While a low concentration of fillers could improve the crystallization of polymer matrix, a high filler loading concentration will suppress the crystallization of matrix and thus leading to a decrease of the thermal conductivity of nanocomposites. [180,181] Recently, the research of Ding et al. [181] showed that the thermal conductivity of GNR@PA6 nanocomposites slightly decreases with the increasing concentration of GNR fillers when GNR concentration was higher than 0.5 wt%. By using X-ray diffraction and differential scanning calorimetry to analyze the morphology of GNR@PA6, Ding et al. unveiled that a low concentration of GNR plays a role in the heterogeneous nucleation to improve the crystallization rate of PA6 and thus enhancing the crystallinity degree, while high concentration of GNR would obstacle the crystallization.

The size of fillers could greatly affect the thermal conductivity of a polymer nanocomposite through influencing the contact area between the fillers and the matrix. Noh et al. [182] showed that larger fillers lead to an obvious thermal conductivity enhancement due to the increased interface conductance. Kim et al. [183] reported a similar result that a larger lateral size and thickness of the GNP may decrease the contact resistance between GNP and polymer matrix to improve the thermal conductivity of nanocomposites. Yu et al. [184] even showed that the filler smaller than a critical size cannot enhance the thermal conductivity because the large interfacial thermal resistance counteract the contribution of the high thermal conductivity of fillers. However, one needs to be careful when choosing the filler size, because other physical properties could be strongly affected when the fillers are too large.

The size of fillers could also greatly affect the thermal conductivity through shortening the mean free path of energy carriers in a nanocomposite incorporated with randomly-distributed fillers. The EMT has been modified to take this effect into account by considering the energy carrier collision cross-section of the fillers and the average distance that the energy carriers can travel inside the fillers [185,186]. The modified EMT could give good predictions compared to the numerical approaches based on the BTE, Monte Carlo simulations and experiments. Further development of modified EMT models has taken into account the spectral phonon properties including MFP, polarizations and wave vectors [187-189].

In short, when no inter-filler networks are formed, the aggregation of fillers and the large interfacial thermal resistance play important roles in the thermal conductivity of polymer nanocomposites [131-135,190]. Here we introduce the methods to reduce the interfacial thermal resistance and to prevent the aggregation of nano-fillers.

### 3.1.1 Methods of enhancing interface conductance

The surface treatment and functionalization of nano-fillers has been widely implemented to enhance the filler-polymer matrix coupling to reduce the interfacial thermal resistance [191-194]. We summarize the major practices according to the type (dimensionality) of fillers.

**1)   1D filler (CNT)**

Although CNT possesses a very high thermal conductivity, the thermal conductivity of CNT@polymer composites is generally lower than 10 $W·m^{-1}·K^{-1}$. The challenge of achieving high effective thermal conductivity primarily originates from the large interfacial thermal resistance between the CNT and the surrounding polymer matrix [195]. The interfacial thermal resistance between CNT and polymer matrix can range from $0.1×10^{-8}$ to $15×10^{-8}$ $m^2·K·W^{-1}$ [196-199]. When heat flows in a CNT composite, the majority of the temperature drop occurs at the interface while the temperature in CNTs essentially remains uniform [199].

Many efforts have already been made to reduce the interfacial thermal resistance between CNTs and polymer matrix through functionalization. MD simulations by Huang et al. [201] showed that the interfacial thermal conductance of functionalized CNTs is enhanced by 100% compared with the pristine CNT-polymer interface, because of the enhanced coupling between CNT and the matrix. Kaur et al. [202] also reported a six-fold reduction of interfacial thermal resistance between matrix and MWCNT arrays by bridging the interface with short, covalently bonded organic molecules. A COOH-functionalization was found to reduce the interfacial thermal resistance, resulting from the modified CNT-matrix interface [203]. Besides functional groups covalently bonded on CNTs, the thermal conductivity could also be enhanced by incorporating silver nanoparticles onto the CNT surface, due to the superior thermal conductivity of silver and the reduced Ag-CNT interfacial resistance

[204].

Although covalent functionalization of nano-fillers could reduce the interfacial thermal resistance, they could also reduce the thermal conductivity of fillers. MD simulations by Ni et al. [205] showed that functionalizing aromatic polymer HLK5 ($C_{22}H_{25}O_3N_3$) onto the CNT can efficiently decrease the interfacial thermal resistance between the CNTs and different types of polymer matrices (PS, epoxy, and PE). However, the thermal conductivity of CNTs can be also decreased by functionalizing HLK5, octane, or hydroxyl on CNTs, as shown in Fig. 18. It shows that a larger functionalized surface of CNTs could lead to a smaller effective thermal conductivity of the polymer composite.

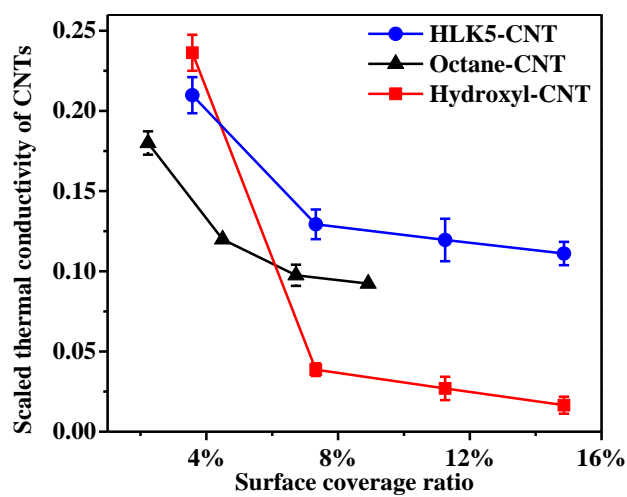

Fig. 18 Thermal conductivity of functionalized CNTs scaled by that of pristine CNT. [205] Copyright 2015 American Chemical Society.

When one end of a molecule group is functionalized on the surface of a CNT and the other end on the polymer matrix, this kind of surface functionalization is also called cross-link. Although the cross-links could improve the thermal coupling between the CNTs and the polymer matrix, they could also suppress the thermal conductivity of CNTs. The thermal conductivity of CNTs can be reduced to half by cross-links compared with that of the freestanding CNTs [206]. The cross-links could also increase the thermal resistance of the matrix. As shown in Fig. 19, the functionalization increases the thermal resistance of the near-interface polymer layer ($R_2$), until $R_2$ saturates when the functionalization degree ($x$) becomes larger than 0.1. The functionalization significantly reduce the interfacial thermal resistance $R_1$ until $x$ reaches 0.15, while a further increase of $x$ will not reduce the interfacial thermal resistance. The result of $R_1 + R_2$ reaches a lowest value at about $x = 0.05$, which suggests that increasing the functionalization degree cannot always lower the total effective thermal resistance [206].

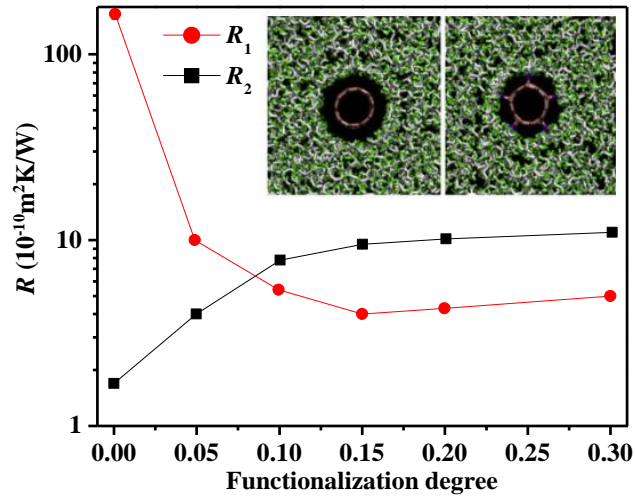

*Fig. 19 Interfacial thermal resistance between CNTs and the PE polymer ($R_1$), and the thermal resistance of the near-interface polymer layer ($R_2$). The functionalization degree is defined as the ratio of the number of functional groups to the number of carbon atoms in the outermost layer of the CNT. Inset shows topologies of the pristine (5,5) SWCNT nanocomposite and the functionalized one with $x = 0.1$. [206] Copyright 2015 Elsevier Ltd.*

**2) 2D fillers (graphene and boron nitride)**

Similar to 1D fillers, the interfacial thermal resistance between 2D fillers and polymer matrix is also the bottleneck for achieving high thermal conductivity in the nanocomposites. For example, the interface conductance between graphene and poly(methyl methacrylate) (PMMA) is as high as that between CNT and polymers, which is about $1.906 \times 10^{-8}$ m²KW$^{-1}$ [207]. To reduce the large interfacial thermal resistance between the 2D-fillers and the matrix, the surface functionalization of 2D fillers, such as graphene, and BN nanosheets has been widely studied. A strong covalent bonding of BN nanosheets on epoxy matrix could result in a higher thermal conductivity [208].

However, surface functionalization of 2D fillers does not always lead to a smaller interfacial thermal resistance. The effect of surface-grafted chains on the thermal conductivity of graphene@polyamide-6.6 nanocomposites has been studied using MD simulations [209]. It turns out that the thermal conductivity perpendicular to the graphene plane is proportional to the grafting density, while the in-plane thermal conductivity of graphene drops sharply as the grafting density increases. There exist an intermediate grafting density for maximal enhancement in the thermal conductivity of nanocomposites, as shown in Fig. 20. Besides the optimal grafting density, there also exists an optimal balance between grafting density and grafting molecular length to obtain the maximum enhancement.

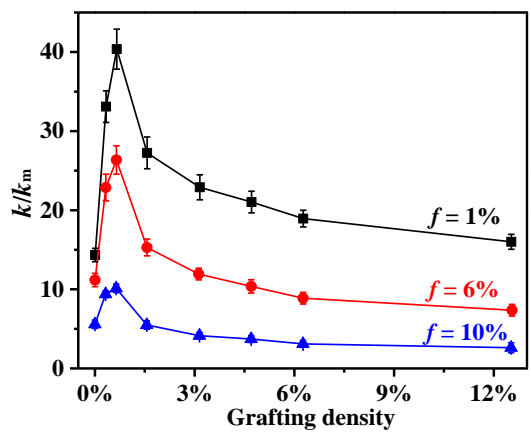

*Fig. 20 Thermal conductivity of nanocomposites scaled by that of the polymer matrix. f is volumetric fraction of graphene fillers. [209] Copyright 2016 American Chemical Society.*

In addition to surface functionalization using organic molecules, inorganic MgO has also been successfully coated on the surface of graphene using a simple chemical precipitation method to enhance the interfacial thermal conductivity, where MgO served as an effective interface to strengthen the interfacial adhesion between the MgO and the epoxy matrix [210]. Non-covalent modification of h-BN nanoparticles with PDA could bring about a higher thermal conductivity of bisphenol-E-cyanate-ester based composites than that without surface functionalization [211]. By depositing silver nanoparticles on the BN nanosheets, Wang et al. [212] showed that the epoxy composite possess a higher thermal conductivity than that with pristine BN nanosheets due to the bridging connections of silver nanoparticles among BN nanosheets.

### 3.1.2 Improvement of filler distribution

In addition to the interfacial thermal resistance between the fillers and the matrix, the thermal conductivity enhancement could also limited by the tendency of filler aggregation. To suppress the filler aggregation, several methods have been developed to improve filler distribution in the past few decades including surface functionalization, addition of dispersant, and special preparation techniques.

CNTs tend to aggregate together because of the strong vdW force and the chemical inertness caused by their unique $sp^2$ bonding [213,214]. To prevent aggregation of CNTs, surface modification of CNTs through covalent or non-covalent approaches have been developed [215,216]. Although chemical functionalization of the CNTs improves the dispersion of CNTs, the covalent functionalization can also distort the structure of CNTs and thus sometimes even reducing the thermal conductivity [217]. Coating CNTs with inorganic materials such as BN or alumina has also been reported to significantly improve the dispersibility of CNTs [218-220]. The thermal conductivities of polyimide (PI) based nanocomposites incorporated with MWCNTs or BN-coated MWCNT (BN-c-MWCNTs) are compared, as shown in Fig.

21 [218]. The thermal conductivity of MWCNT@PI decreases with the increasing concentration of MWCNTs when its concentration is larger than 1 wt%, while the thermal conductivity of BN-c-MWCNT@PI keeps increasing with the increase of BN-c-MWCNT concentrations. Such different results originate from the different dispersibility of MWCNTs and BN-c-MWCNT in the PI matrix. The dispersibility of MWCNTs is poor, and the aggregation happens when its concentration is higher than 1 wt%, which increase the interfacial thermal resistance between MWCNTs and polyimide. For BN-c-MWCNT@PI composites, the fillers do not aggregate because of the BN coating. Similar phenomenon was also observed in Ref. [221] that by coating PVP on the surface of silver nanowire fillers could hinder their aggregation in a silver-nanowire@epoxy nanocomposite.

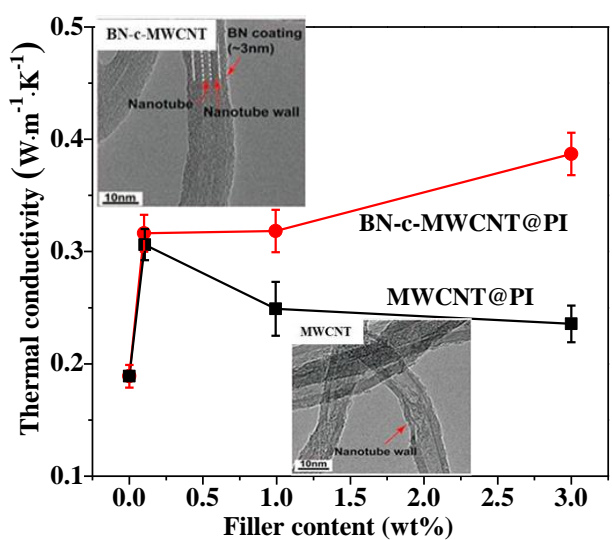

*Fig. 21 Thermal conductivity of MWCNT@PI and BN-c-MWCNT@PI. Insets are the transmission electron microscope images of MWCNTs and BN-c-MWCNTs. [218] Copyright 2014 Royal Society of Chemistry.*

Unfortunately, it remains very challenging to simultaneously achieve strong filler-matrix coupling and low aggregation. Surface functionalization for suppressing filler aggregations usually leads to a weak filler-matrix coupling, making it difficult to achieve high thermal conductivity [222]. In addition to the surface treatment and functionalization, dispersant is another choice to prevent aggregation of fillers. graphene oxide (GO) appears to be a good dispersant, as it has similar lattice structure to graphene and could also lead to a stable dispersion of CNTs in aqueous media through the π-π stacking interaction between GO and CNTs [223,224]. Clay could also promote a uniform distribution of CNTs to form a percolated network structure in a polymer matrix [224,225]. Polyvinylpyrrolidone (PVP) is helpful to improve the dispersion of CNTs and induce denser CNT network structure in poly(vinylidene fluoride) PVDF matrix [226].

Some special preparation process could also improve filler distribution. The thermal conductivity of the BN@PE composites has been improved after multistage stretching extrusion process, because the aggregation of BNs were reduced due to the strong shear field [227]. On the contrary, rheological studies showed that the shear could also induce an aggregation of CNTs, and CNTs with a longer aspect ratio possess a larger shear-induced aggregation [228]. The shear-induced distribution of fillers depends on not only the filler but also on the processing details. Some practical and feasible preparation are still needed to improve the distribution of CNT and other fillers in a polymer matrix.

The orientation of fillers with large aspect ratio is another crucial factor determining the effective thermal conductivity of nanocomposites. Similar to the polymers, the alignment of filler orientation can be improved by tension [229,230], templating [231,232] and electrospinning [233]. For example, Datsyuk et al. [233] prepared a high thermally conductive CNT@polybenzimidazole nanocomposite by electrospinning. The in-plane thermal conductivity of the composite reaches 18 $W \cdot m^{-1} \cdot K^{-1}$ at 1.94 wt% CNT concentration, 50 times larger than the polymer matrix. This high thermal conductivity is attributed to the excellent CNT alignment in the polymers.

Electromagnetic field or electric field could be applied to align fillers. For example, Lin et al. [234] utilized an external magnetic field to align hexagonal boron nitride (h-BN) to obtain a high thermal conductivity along the alignment direction. Cho et al. [235] used a high direct current electric field to enhanced thermal conductivity in the out-of-plane direction of the BN@polysiloxane nanocomposites. The ice templating self-assembly strategy is a proven effective method to form well-aligned fillers along the ice-growth direction [236 - 238]. The thermal conductivity of BN@epoxy-resin nanocomposites prepared with this strategy is as high as 4.42 $W \cdot m^{-1} \cdot K^{-1}$, much higher than that of nanocomposites filled with randomly distributed BN which is less than 1.81 $W \cdot m^{-1} \cdot K^{-1}$ [239].

**3.2 Nanocomposites with inter-filler network**

It is natural to expect that an inter-connected network can be formed when the filler concentration is higher than a critical value, i.e. the percolation threshold [240]. Although a single kind of fillers can form a 3D network, there still remains lots of voids that might not be filled tightly by the polymer matrix. The introduction of another kind of fillers is beneficial to fill the vacancies and to further enhance the effective thermal conductivity of nanocomposites. Incorporating with only a small amount of hybrid fillers could lead to an equally increased thermal conductivity with respect to that filled with single fillers. However, the thermal conductivity of nanocomposites even with the inter-filler network is still too low compared to many high thermal conductivity inorganic materials, due to the large inter-filler thermal contact resistance. Recently, porous 3D fillers, such as carbon foam and graphene foam, have drawn great interest because of their intrinsic 3D network structure with no thermal contact resistance. In Sections 3.2.1-3.2.3, we discuss three kinds of nanocomposites with 3D networks.

**3.2.1 Single fillers**

The filler network could be greatly influenced by the dimensions and shapes of individual fillers, and thus the thermal conductivity of polymer nanocomposites [241-244]. In a nanocomposite filled with a single kind of fillers, the influence of filler dimensionality on the thermal conductivity ($k$) generally follows the trend of $k_{2D} > k_{1D} > k_{0D}$, where 2D, 1D and 0D stand for nanocomposites incorporated with 2-dimentional, 1-dimentioanl and 0-dimensional fillers respectively, because the total contact area per volume increases with the increase of filler dimensionality [245-247]. Furthermore, the contact area can have more influence on the thermal conductivity of fillers with lower dimensionality than larger dimensionality. The defects such as kinks and twists formed in each individual CNT could lead to a reduction of its thermal conductivity, while defects in 2D fillers have no such important influence. For example, platelet-like 2D fillers are preferred for enhancing the thermal conductivity than a 1D nanowires and nanotubes, because of a lower thermal contact resistance of a surface than a point contact resistance. Among 2D GNP, 1D CNT, and 0D super-fullerene, the GNP-enhanced PVDF composites possess the highest thermal conductivity (2.06 W·m$^{-1}$·K$^{-1}$), which is about 10-folds enhancement compared to that of the pristine PVDF [248]. The GNP@epoxy shows a thermal conductivity of 0.47 W·m$^{-1}$·K$^{-1}$ with an increase about 126.4 % compared to that of the neat epoxy, while MWCNT@epoxy shows a thermal conductivity of 0.33 W·m$^{-1}$·K$^{-1}$, an increase for only about 60 % [249].

  The shape of fillers also plays an important role. Among different shapes of 2D fillers, the highest thermal conductivity enhancement is found to be using the prolate ellipsoids compared to other shapes like oblate ellipsoids and spheres, because it is easier to form a conductive network with prolate ellipsoids [241-243]. Although 2D fillers are better than 1D fillers for enhancing the thermal conductivity because of the face-contact between 2D fillers, the 1D filler is more apt to form a heat-transport inter-filler network at a lower concentration which in turn might enhance the thermal conductivity due to their large aspect ratio. The percolation threshold for 2D and 0D fillers is much higher than that of 1D fillers because of different aspect ratios where the theoretical percolation threshold for heat transport is proportional to the reciprocal of aspect ratio [250]. Due to the high aspect ratio of CNTs (up to 10$^3$-10$^4$), it is possible to establish percolation paths at a low CNT concentrations, usually lower than 1 vol%, to enhance the thermal conductivity. It has been shown that enlarging the aspect ratio of CNT to reduce the number of contacts required to form a percolating network could improve the thermal conductivity of the composite, [158] and the thermal conductivity of CNT@polymer composites can be effectively controlled by adjusting the length of the CNT fillers [251]. On the other hand, the percolation threshold for 2D and 0D fillers are usually much higher than 1 vol%. For example, the percolation threshold of the thermal conductivity in PE based nanocomposites incorporated with graphite powder is as high as 10 vol% [252].

  The research carried out by Zhao et al. [253] is introduced here to illustrate the thermal conductivity change with a percolation phenomenon when an inter-filler network is formed, as shown in Fig. 22. Three different fillers, raw carbon fiber (raw-CF), copper-coated CF (Cu-CF) and (3-mercaptopropyl)trimethoxysilane-Cu-CF (M–Cu-CF), were incorporated in silicone rubber (SR) matrix to enhance the thermal conductivity. These polymer composites all exhibited an enhanced thermal conductivity with a percolation threshold at the filler concentration of about 1 wt %. The larger thermal conductivity of nanocomposites filled with Cu-CF and M–Cu-CF than that with CF is due to the fiber surface treatment. When the filler loading is less than the percolation thresholds, the thermal conductivities increase slowly, because of the large interfacial thermal resistance between fillers and matrix. When the filler loadings increased up to about 4.0 wt%, there is a dramatic increase of the thermal conductivity resulting from the networks. Nevertheless, the thermal conductivities increase slowly again when the filler concentration becomes larger than 4.0 wt%, because the thermal conductive paths tend to saturate. Besides the widely observed percolating behavior of polymer nanocomposites incorporated with 1D-fillers where the increase of thermal conductivity follows a critical power law [253-255], it has also been widely observed in many experiments that a small amount of 2D fillers filled in a polymer matrix could result in a significant enhancement of the thermal conductivity [256, 257].

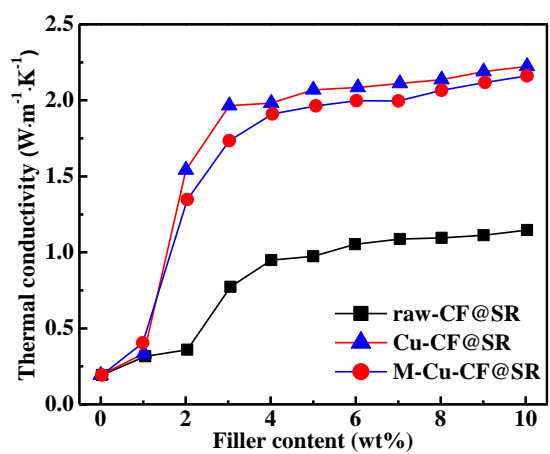

*Fig. 22 Experimental thermal conductivity of raw-CF@SR, Cu-CF@SR, and M–Cu-CF@SR nanocomposites. [253] Copyright 2016 Springer International Publishing AG.*

Among 0D, 1D and 2D fillers, the nanocomposites incorporated with 1D fillers are most well-studied. The heat conduction through the network in a nanocomposite filled with 1D fillers is influenced by many factors such as inter-filler thermal contact resistance, intrinsic thermal resistance of the fillers, volumetric fraction, aspect ratio of fillers, and orientation distribution of fillers. Most recently, a general thermal conductivity model for nanocomposites with 1D-filler network has been developed to include the effects of inter-filler thermal contact resistance, intrinsic thermal resistance of fillers, volume fraction of fillers, and orientation distribution of fillers [258]. To describe the competing effect of the inter-filler thermal contact resistance and the filler intrinsic thermal resistance, Zhao et al. [258] defined a dimensionless

Biot number to describe this competing effect,

$$Bi_T = \frac{\langle N_c \rangle hL}{k_0 A_0} \quad (1)$$

where $\langle N_c \rangle$ is the average number of contact junctions for a single fiber with its neighboring fibers, $h$ is the thermal conductance of the inter-filler contact, $k_0$, $L$ and $A_0$ is respectively the thermal conductivity, length and cross-sectional area of the 1D fillers, as shown in Fig. 23(a). The thermal conductivity of polymer nanocomposites with filler networks could then be derived as, [258]

$$\frac{k}{k_0} = n_s \frac{Bi_T}{2\langle \cos\theta \rangle L/\langle H \rangle + Bi_T} \langle \cos\theta \rangle \quad (2),$$

where $n_s$ is number of fillers across an arbitrary cross-section with unit area, $\langle H \rangle$ is the average center-to-center distance of two connected fillers in the heat transfer direction, $\theta$ is the angle between the axial direction of the filler and the direction of heat transfer.

It can be shown from Eq. 2 that a larger $\theta$ could lead to a higher thermal conductivity of polymer nanocomposites. Similarly, the filler-orientation-improvement methods discussed in Section 3.1.2 may be also useful for improving the thermal conductivity of nanocomposites with filler networks, although such kind of studies has yet been carried out. Reducing the inter-filler thermal contact resistance is helpful in increasing the effective thermal conductivity. The increasing trend of thermal conductivity with increasing volume fraction of fillers greatly depends on the $Bi_T$, as shown in Fig. 23(b). For the nanocomposite filled with CNTs, there is a large inter-filler thermal contact resistance ($Bi_T \ll 1$) and the thermal conductivity is sensitive to the volumetric fraction of fillers.

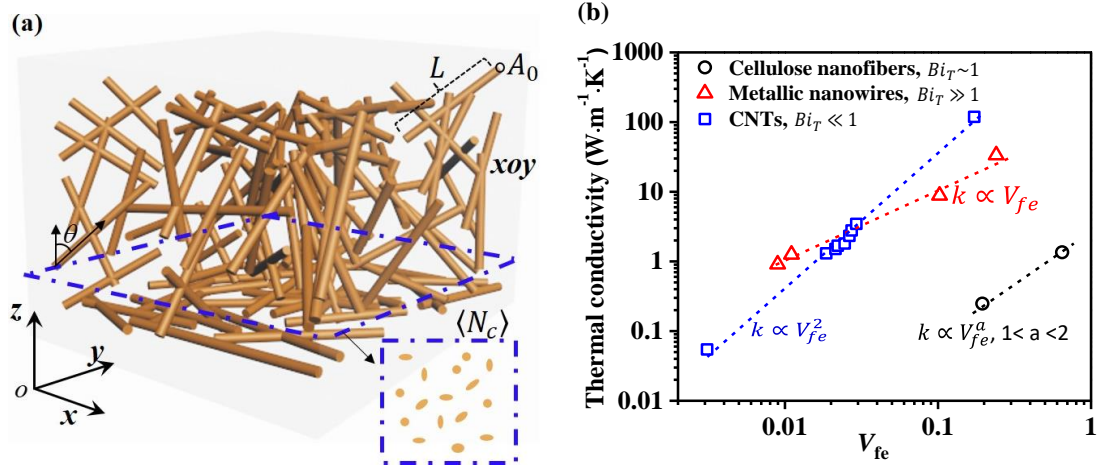

*Fig. 23 Thermal transport in a nanocomposite filled with 1D fillers: (a) Schematic of a nanofiber network; (b) thermal conductivity versus volume fraction ($V_{fe}$). Lines showing model results, while circular, square and triangle dots stand for experimental results. [258] Copyright 2017 American Institute of Physics.*

### 3.2.2 Hybrid fillers

Although a single kind of fillers can form a 3D network, there still remains lots of voids that can hardly be filled by the polymer matrix. The introduction of another kind of fillers is beneficial to fill the voids and to further enhance the effective thermal conductivity of nanocomposites. There is always a preferable concentration ratio between different constituent fillers to enhance the thermal conductivity [259-261]. By studying the thermal conductivity of polycarbonate (PC) based nanocomposites filled with MWCNTs and EG, Zhang et al. [262] showed that the maximum thermal conductivity could be achieved with loading weight ratio of 9:1 between the EG/MWCNTs, and this maximum thermal conductivity is higher than that of the nanocomposite with either single kind of fillers. However, there yet exists a theory to find the optimal concentration ratio. More studies are still needed to probe the relationship between the best concentration ratio and the geometric properties of hybrid fillers like sizes and dimensionality. It should be also noted that the optimal concentration ratio of hybrid fillers not always ensure a maximum thermal conductivity, whereas a much higher concentration of fillers may destroy the synergistic effect of hybrid fillers [263].

According to different structures of filler networks, the polymer nanocomposites incorporated with hybrid fillers can be categorized into four cases: (1) both of two fillers uniformly disperse in the nanocomposite without a network because of a low filler concentration (without network); (2) one of the fillers forms a network while another filler remains uniformly dispersed (single network, no synergistic effect); (3) while one of hybrid fillers distributes uniformly, another filler bridge them together to form a network (single network with synergistic effect); (4) both kinds of fillers form network, and double networks are cross-linked (double network with synergistic effect). Here we focus on the synergistic effects, i.e., cases 3 and 4, which might be inspiring to develop new methods to realize the synergetic enhancement of thermal conductivity.

**1) Single network with synergistic effect**

The single filler network with different hybrid fillers are schematically shown in Fig. 24(a)-(c), which are formed with 0D+1D hybrid fillers, 0D+2D hybrid fillers and 1D+2D hybrid fillers, respectively.

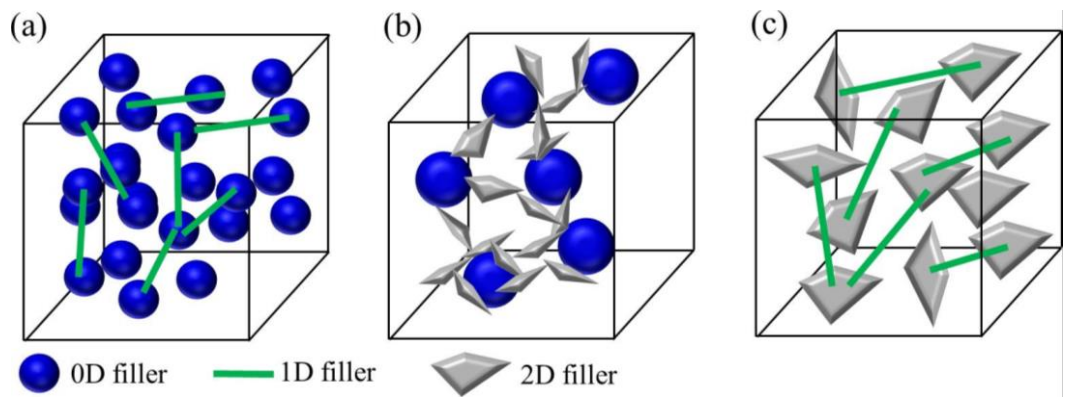

*Fig. 24 Schematic diagrams of single network formed with different hybrid fillers: (a) 0D+1D fillers; (b) 0D+2D fillers; (c) 1D+2D fillers*

**(a) With 0D+1D hybrid fillers**

In the nanocomposite incorporated with 0D+1D hybrid fillers, a single network with synergistic effect could be formed with one of hybrid fillers distributed uniformly and another bridging them together. The 0D+1D hybrid fillers contribute considerably to the formation of a more efficient percolating network for the thermal conduction and thus improving the thermal conductivity, compared to that of single fillers. For example, the thermal conductivity of PVDF+PS polymer blend based nanocomposites incorporated with MWCNTs+SiC hybrid fillers could be 1.85 W·m$^{-1}$·K$^{-1}$ [264], as that shown in Fig. 25. However, the composite incorporated with only one kind of fillers exhibits a thermal conductivity of only 0.40 W·m$^{-1}$·K$^{-1}$ for 2.9 vol% MWCNTs and 0.98 W·m$^{-1}$·K$^{-1}$ for 11.4 vol% SiC. There is an obvious synergistic effect between MWCNTs and SiC on the thermal conductivity. In this single network with synergistic effect, the MWCNTs act as heat conducting bridges among the SiC nanoparticles, while SiC could separate MWCNTs to prevent the aggregation, as that shown in the inset of Fig. 25.

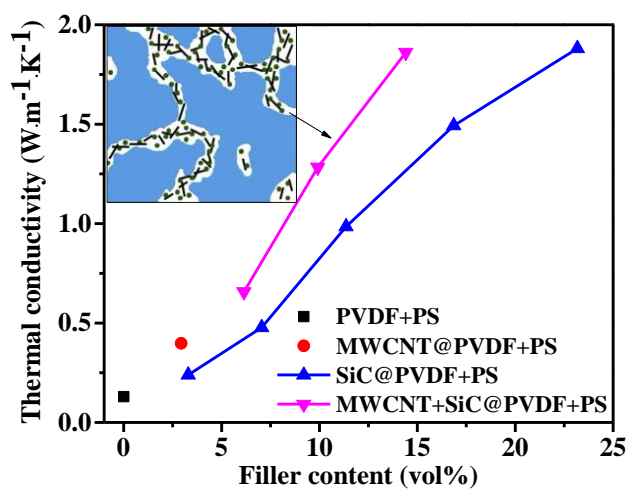

*Fig. 25 Thermal conductivity of PVDF+PS blends filled with different types of fillers. The concentration of MWCNTs is maintained as 2.9 vol % for all composites. The data of the SiC@PVDF+PS are obtained from Ref. [265]. The inset schematically illustrates the distribution of MWCNTs and SiC in MWCNT+SiC@PVDF+PS nanocomposite, where the white part represents the PVDF, the blue part PS, the black lines MWCNTs, and the green circles SiC. [264] Copyright 2013 Elsevier Ltd.*

**(b) With 0D+2D hybrid fillers**

The thermal conductivity of poly(3-hydroxylbutyrate) (PHB) based nanocomposites with a total filler concentration of 50 wt % but different concentration ratio between BN and Al$_2$O$_3$ was studied by Li et al. [266]. Fig. *26* shows that the maximum thermal conductivity of BN+Al$_2$O$_3$@PHB is 31 % higher than that of the BN@PHB composites and 196 % higher than that of the Al$_2$O$_3$@PHB composites, respectively. This suggests a synergistic effect of hybrid fillers on the thermal conductivity in BN+Al$_2$O$_3$@PHB. Fig. 26(b) shows that with a small volumetric fraction of Al$_2$O$_3$ particles, some BN nano-sheets tend to align along the Al$_2$O$_3$ surfaces to form a network with a greatly increased contact area. The thermal conductivity is therefore improved with the concentration increase of Al$_2$O$_3$ particles. When the concentration of Al$_2$O$_3$ becomes larger than a certain value, the Al$_2$O$_3$ particles form a network rather than BN which would disperse in gaps among the Al$_2$O$_3$ particles, thus a decrease of thermal conductivity occurs. The optimal ratio of BN and Al$_2$O$_3$ for obtaining a maximum thermal conductivity is 43:7 (43 wt% BN and 7 wt% Al$_2$O$_3$), where the maximum thermal conductivity is about 1.79 W·m$^{-1}$·K$^{-1}$.

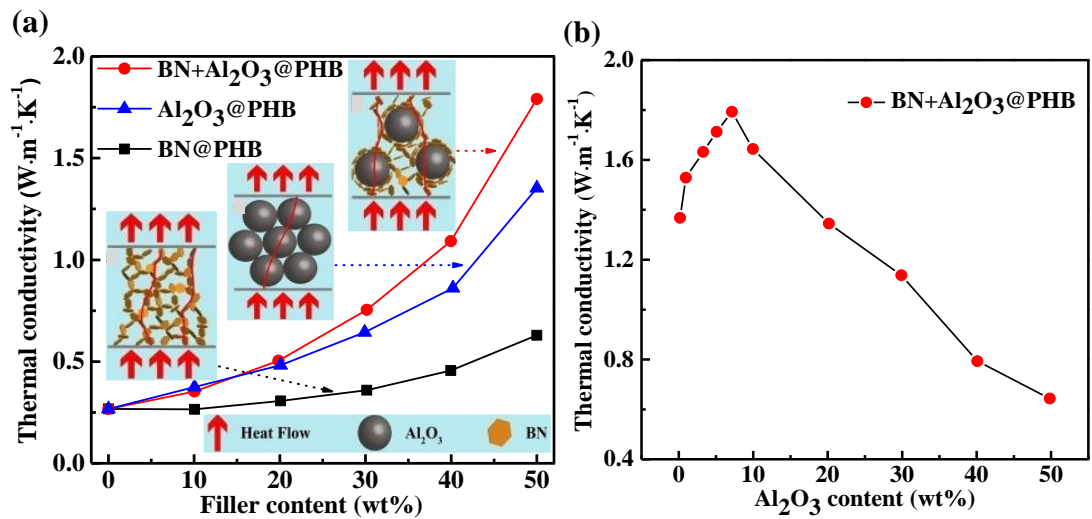

*Fig. 26 Experimental thermal conductivity of PHB composites: (a) synergistic effect; (b) effect of hybrid filler concentration ratio, with a total filler concentration of 50 wt%. [266] Copyright 2017 Elsevier Ltd.*

**(c) With 1D+2D hybrid fillers**

Similar synergistic enhancement effect could be achieved with 1D+2D hybrid fillers. In such kinds of nanocomposites, the CNT is usually applied as the 1D filler, and the BN or GNP is commonly used as 2D fillers. The high rigidity of GNPs can impede their bending in a high volume fraction and thus preserves their high aspect ratio for providing more islands to be connected by CNTs. A rigid 2D filler in the hybrid fillers could also stop bending and coiling of CNTs in the nanocomposite, and thus maintaining the aspect ratio and high thermal conductivity of CNTs.

He et al. [267] studied the enhancement of polymer-bonded explosives (PBX) using the hybrid fillers of 1D CNTs and 2D GNPs. The thermal conductivity of PBX based nanocomposites with a filling ratio of 1.31 vol% comprising of 10 vol% GNPs and 90 vol% CNTs is about 1.4 W·m$^{-1}$·K$^{-1}$, which is more than twice that of PBX with pure GNPs or CNTs. This high thermal conductivity is due to the synergistic effect of hybrid fillers when GNPs and CNTs could lead to the formation of a more efficient 3D percolating network. The synergetic effect of cyanate ester (CE) resin based nanocomposites is also studied, which is incorporated with hybrid fillers, including dodecylamine-modified GNPs (da-GNPs) and γ-aminopropyl-triethoxysilane-treated MWCNTs (f-MWCNTs) [268], as shown in Fig. 27(a). The composite with 3 wt% hybrid filler exhibits a 185 % increase in thermal conductivity compared with that of CE resin matrix, while composites with individual da-GNPs and f-MWCNTs exhibits an increase of 158 and 108 %, respectively. The concentration ratio of hybrid fillers also exerts a large effect on the thermal conductivity, when 3:1 is the optimal concentration ratio of da-GNPs and f-MWCNTs for maximizing the thermal conductivity.

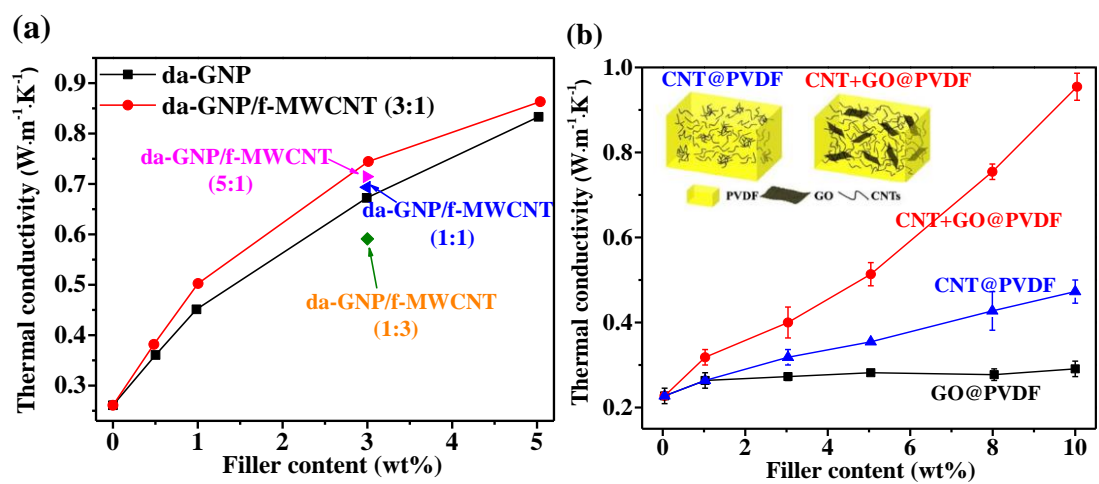

*Fig. 27 Experimental thermal conductivity of polymer nanocomposites incorporated with hybrid fillers: (a) cyanate ester resin based nanocomposites filled with da-GNP/f-MWCNT hybrid fillers. [268] Copyright 2014 John Wiley and Sons; (b) PVDF nanocomposites with CNT+GO hybrid fillers, the filler concentration of GO is kept at a constant 1wt%. Inset shows the different dispersion states of CNTs in the CNT@PVDF and the CNT+GO@PVDF composites. [273] Copyright 2015 Elsevier Ltd.*

The thermal conductivity of composites filled with hybrid fillers of BN and MWCNTs is widely studied, and a synergistic improvement in the thermal conductivity was also observed [269-272]. Xiao et al. [270] reported that with a small concentration of CNTs incorporated in BN@PVDF composites via melt blending method, the thermal conductivity of the composite is much higher than that of BN@PVDF composites at the same BN concentration. In addition to the widely studied hybrid fillers of CNT and BN or GNP, Zhang et al. [273] also found a greatly improved thermal conductivity in the CNTs+GO@PVDF nanocomposites with respect to that of CNT@PVDF nanocomposites at the same CNT concentration, as shown in Fig. 27(b). It was demonstrated that the introduction of GO is beneficial to the dispersion of CNTs and the formation of denser CNT+GO networks in the PVDF matrix. Appling same hybrid fillers, a similar synergetic effect was also realized for PP polymer based nanocomposites where the highest thermal conductivity was reported to be ~ 0.35 W·m$^{-1}$·K$^{-1}$ [274].

2) **Double networks with synergistic effect**

Hybrid fillers can form interconnecting double networks, especially with large aspect ratio 1-D and 2-D nano-fillers. The effect of the double networks with hybrid 1D+2D fillers on the thermal conductivity can be summarized as [275]: (1) extremely low thermal contact resistance achieved by overlaping interconnections within 2D fillers ; and (2) synergistic effect between 1D-filler network and 2D-filler network based on the bridging effect as well as increasing the network density.

For polymer nanocomposites incorporated with CNT and EG hybrid fillers, double networks are usually formed at a high concentration of fillers, as schematically shown in Fig. 28(a). Two interpenetrating networks are formed instead of just an EG network wrapping CNTs. Such double networks provide more efficient heat conducting paths, and a sharp increase of thermal conductivity is expected at just above the percolation concentration even with only a small amount of CNTs for HDPE/15EG/xCNT and HDPE/20EG/xCNT nanocomposites [276], as shown in Fig. 28(b). The thermal conductivity of HDPE/10EG/xCNTs increases linearly with the concentration increase of CNT fillers along the same line by adding the same amount of EG, which suggests that no synergistic effect exists in this system but only a mix role is obeyed. Wu et al. [277] also reported a sharp increase of the thermal conductivity in PP polymer based nanocomposites via the formation of double percolated filler network with small-sized MWCNT network located within loose large-sized EG network. Their results suggested that the formation of double networks could effectively reduce the inter-filler thermal contact resistance and thus significantly increasing the effective thermal conductivity of nanocomposites.

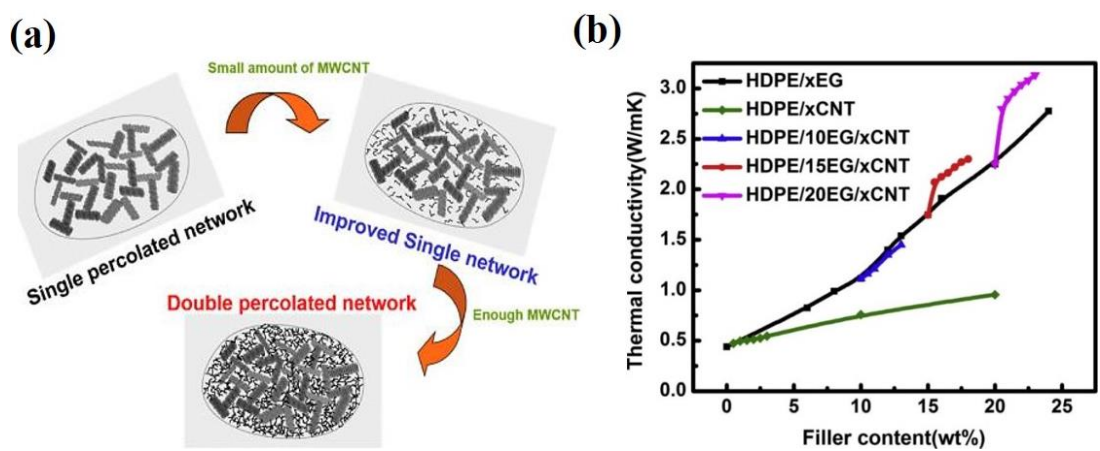

*Fig. 28 CNT+EG@HDPE nanocomposite: (a) Schematic illustrating double percolated networks constructed by the hybrid fillers EG and MWCNT; (b) Thermal conductivity of CNT+EG@HDPE nanocomposites with the concentration increase of CNT while maintaining the EG concentration at 10, 15 and 20 wt% respectively. The*

*10EG/xCNT, 15EG/xCNT and 20EG/xCNT stand for increasing the concentration of CNT while maintaining the EG concentration at 10, 15 and 20 wt% respectively. [277] Copyright 2016 Elsevier Ltd.*

A remarkable synergetic effect with cross-linked double networks on the thermal conductivity was also observed in the epoxy-resin based nanocomposites incorporating with GNP and MWCNTs hybrid fillers, [278] as shown in Fig. 29(a). The thermal conductivity of nanocomposites with 20 vol% CNTs and 20 vol% GNPs could be as high as 6.3 W·m$^{-1}$·K$^{-1}$, which is much higher than that of the composites with 50 vol% CNTs or 50 vol% GNPs alone. The maximum thermal conductivity is as high as 7.30 W·m$^{-1}$·K$^{-1}$, which is about 38 times that of epoxy-resin matrix. This maximum thermal conductivity is much higher than that when a single network is formed with same hybrid fillers as shown in Fig. 27, which suggests that the cross-linked double networks is much better than a single network to enhance the thermal conductivity. However, it should be noted that a much higher filling ratio is usually required to form a cross-linked double network compared with that of a single network. To reveal the influence of the filling ratio on the synergistic effect of hybrid fillers to enhance thermal conductivity, the strength of the synergistic effect is defined as $(k_{HYB} - k_{GNP})/k_{GNP}$ by Huang et al. [278], where $k_{HYB}$ and $k_{GNP}$ is the thermal conductivity of nanocomposites filled with hybrid fillers and only GNP fillers, respectively. The strength of the synergistic effect is shown in Fig. 29(b). It shows that the synergistic enhancement of thermal conductivities happens in the filler concentration ranging from 10 to 50 vol%, and the synergistic effect is more remarkable at a high filler concentration.

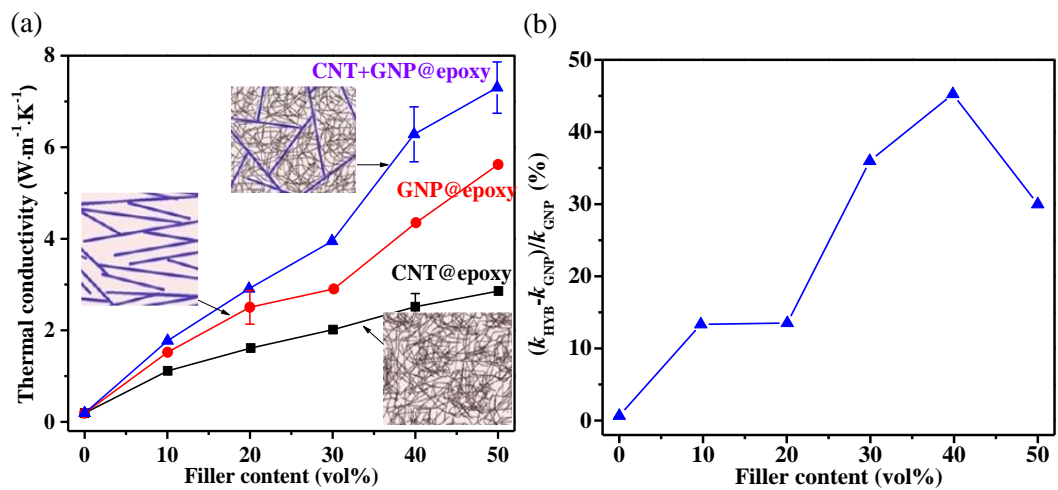

*Fig. 29 Thermal conductivities of epoxy based nanocomposites: (a) synergistic effect of cross-linked double networks with the hybrid filler concentration ratio of 1:1 for CNT+GNP@epoxy nanocomposites; (b) strength of the synergistic effect. [278] Copyright 2012 American Chemical Society.*

### 3.2.3 3D foam fillers

The self-supported 3D fillers, such as 3D graphene foam (GF) and carbon foam (CF) can overcome the shortcoming of filler aggregation during the manufacturing process of nanocomposite, and thus providing a more stable 3D thermal transport network to enhance the thermal conductivity [279,-281]. The worm-like 3D EG is formed when the pristine graphite is intercalated with a variety of inserting agents at high temperature. The nanocomposites incorporated with 3D EG fillers can be prepared with a melt-blending method due to its good affinity with polymers and the intumescent nature [282-287]. The thermal conductivity of PA6 based nanocomposite with an EG concentration of 60 wt% could be as high as 21.05 $W·m^{-1}·K^{-1}$, approximately 72 times higher than that of the matrix [286]. Incorporated with ultrathin-graphite foams at a low filling ratio of about 0.8-1.2 vol%, the wax based nanocomposite could possess a thermal conductivity of about 3.5 $W·m^{-1}·K^{-1}$, which is 18 times that of the matrix [288].

To illustrate the dependence of the thermal conductivity on 3D-foam structure, microstructures of GF and CF are respectively shown in Fig. 30(a) and (b), [279,281] and the thermal conductivity between GF and CF are compared in Fig. 30(c). It shows that the thermal conductivity depends exponentially on the mass density for both GF and CF, because there could be more walls for a larger density. The GF with nanoscale strut-wall is more effective in enhancing thermal conductivity than that with microscale strut-walls, because there could be more heat transfer routes in GF with nanoscale walls at a given GF volume [279]. It is preferable to increase the volume fraction of GF through reducing the pore size without increasing the strut wall thickness [279]. By backfilling PMMA into the pores of GFs, 3D thermal conductive paths could be formed in the nanocomposite at an extremely low concentration of GF, and thus a significant increases of thermal conductivity could be achieved (0.35-0.70 $W·m^{-1}·K^{-1}$) [207]. Li et al. [289] also reported that the thermal conductivity of polyamide-6 (PA6) based nanocomposite filled with 3D GF could be improved by 300 % to 0.847 $W·m^{-1}·K^{-1}$ at a GF loading of 2.0 wt%.

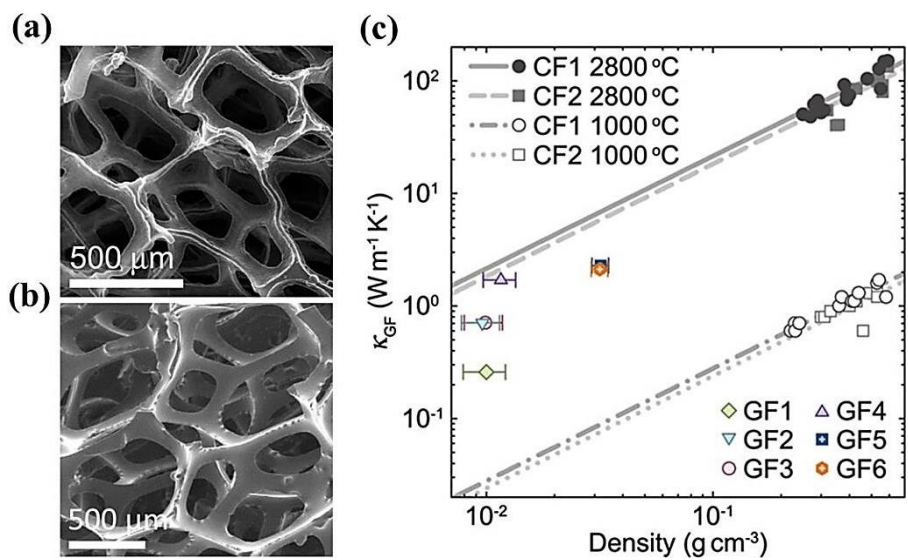

*Fig. 30 Structures and thermal conductivities of GF and CF: (a) scanning electron*

*microscopy images of a GF [279]; (b) scanning electron microscopy images of a CF [281]; (c) thermal conductivity comparison between GF and CF [281]. CF1 and CF2 are respectively derived from Mitsubishi ARA24 and Conoco Dry Mesophase. 1000 and 2800 °C is the graphitization temperature. GF1-6 signifies GF prepared with different methods. Linear fits to the density dependence of the thermal conductivity. (c) is reprint here with the permission of Copyright 2000 from Elsevier Ltd.*

The associated high electrical conductivity of the GF and CF sometimes limits their applications in some fields that requires electrical insulation. The 3D boron-nitride nanosheets (3D-BNNs), which possesses a high thermal conductivity but a quite low electrical conductivity can be used as an alternative [290-292]. The 3D-BNNs aerogel exhibits a much better ability to enhance the thermal conductivity of epoxy resins based nanocomposites, and the thermal conductivity is about 181 % higher than that incorporated with BN, due to the unique 3D structure of 3D-BNNs aerogels which has a smaller inter-BNNs interfacial thermal resistance [290]. By assembling of BNNs on a 3D cellulose skeleton, Chen et al. fabricated a cellulose nanofiber-supported 3D interconnected BNNs (3D-C-BNNs) [293]. This 3D-C-BNNs also has an ultra-high thermal conductivity about 14 times that of matrix at a low concentration of BNNs (i.e., 9.6 vol%), which is about 3.13 $W·m^{-1}·K^{-1}$.

## 4 Summary and outlook

In this review, we focused on the influencing factors and their underlying physical mechanisms for tailoring thermal conductivity of polymers and polymer nanocomposites. The main research progress over the last two decades can be summarized as:

**1)** Improving crystallinity or chain alignment of polymers usually enhances polymer thermal conductivity. Selecting appropriate polymer species with special chain structures is crucial to further increase the thermal conductivity. With the influence of the chain structure widely studied, it is clear that a larger stretching and bending strength of backbone bond leads to a higher thermal conductivity, while small dihedral-angle strength can significantly reduce the thermal conductivity. Besides bond strength of the backbone, a larger weight or number density of side chains give a lower thermal conductivity, while the effect of the side chain length may depend on the morphology of side chains.

**2)** The thermal conductivity of pristine polymers could be increased by enhancing inter-chain coupling, such as through H-bonds and covalent cross-links. The enhancement effect is attributed to the following reasons: more inter-chain "thermal bridges", suppress rotation of polymer chains, and the extended chain coil. On the contrary, it is also widely reported that the inter-chain coupling could decrease the thermal conductivity, because they may cause phonon scatterings and suppress the contribution of acoustic phonon modes. More studies are still needed to further clarify the concurrent effects.

**3)** When nano-filler aggregates in the polymer composite, the thermal conductivity could exhibit a non-monotonic increasing trend as a function of the filler concentration, which is controlled by the aggregation of fillers and the filler-matrix interfacial thermal resistance. The thermal conductivity increases at low concentration due to the higher thermal conductivity of nano-fillers. The aggregation of fillers reduces the thermal conductivity at an intermediate concentration. If the concentration is further increased, a second rise of the thermal conductivity may occur because the clusters formed by aggregation of fillers could contact with each other to form heat transport networks.

**4)** To prevent the aggregation of nano-fillers, nano-fillers are often functionalized. However, surface functionalization for hindering filler aggregations could also lead to a larger interfacial thermal resistance. It is highly desirable to simultaneously reduce the interfacial thermal resistance and suppress aggregation of nano-fillers.

**5)** For nanocomposites filled with uniformly distributed fillers, the thermal conductivity first increases slowly until the filler concentration is larger than the percolation threshold, then grows rapidly and finally converges to a constant value when the filler network saturates. While it is still a great challenge to form a 3D network with a small addition of single fillers, hybrid fillers are usually applied. Incorporating with hybrid fillers, single or double networks can be formed. Contrasting to the composite with single networks, the composite with double filler networks could have a much higher thermal conductivity, due to the reduced inter-filler thermal conduct resistance and the bridging effect between double networks. For hybrid fillers, there is always an optimal concentration ratio between two fillers for enhancing the thermal conductivity. Nevertheless, this ratio is still difficult to be determined except by experiments, and it is still an open question to completely understand how to modulate the effect brought about by hybrid fillers.

**6)** The 3D fillers could be more effective than hybrid fillers to enhance the thermal conductivity because of their intrinsic 3D network without thermal contact resistance, such as CF, GF, EG and 3D-BNNS aerogel. The thermal conductivity of 3D fillers depends exponentially on the mass density, because larger density could lead to more strut-walls.

Although significant progress has been made over the last two decades in enhancing the thermal conductivity of polymers and polymer nanocomposites, there is still a lot of work need to be done, given the technical importance of nanocomposites. Here we list a few directions that might be worthwhile for exploration:

**1)** The enhancement of the intrinsic thermal conductivity of polymers is only experimentally realized by enhancing chain alignment and inter-chain coupling with polymer blends. Furthermore, only the thermal conductivity along the alignment direction is enhanced while an isotropic high thermal conductivity material is commonly desirable, which might limit the applications. Blending polymers could only obtain a slightly enhanced thermal conductivity which is usually lower than 0.5 $W \cdot m^{-1} \cdot K^{-1}$ and this method is difficult to be adopted because of complex synthesis conditions. To further enhance the thermal conductivity, appropriate polymer species should be selected based on the theoretical studies before carrying out the chain alignment and polymer blending.

**2)** Although incorporating fillers in a polymer could improve the thermal conductivity, it is still a challenge to fabricate polymer-based nanocomposites with high thermal conductivity because of some well-known difficulties such as filler aggregation and large inter-filler thermal contact resistance. 3D fillers, such as graphene foams and carbon foams, should attract more attentions in the future, because of their intrinsic 3D network structure without thermal contact resistances, which could be more effective than other kinds of fillers, such as 1D CNTs and 2D graphene nanoplates.

**3)** Combining the demonstrated methods emerged for the thermal conductivity enhancement in both polymers and polymer nanocomposites together may provide some routes to obtain even higher thermal conductivity, for example, mechanically stretching polymer blends based nanocomposites filled with graphene foams.

**4)** To theoretically understand the thermal transport in polymer nanocomposites, many models have been developed to evaluate the effect of the size, shape, intrinsic thermal conductivity and dispersion of fillers, but few works has explored the effectiveness of hybrid fillers [294]. The thermal conductivity mechanisms in composites near percolation have yet to be elucidated [295].

**Acknowledgments**
CH acknowledges the financial support by the National Natural Science Foundation of China (No. 51406224). RY acknowledges the financial support by National Science Foundation (Grant No. 0846561, 1512776), DOD (Grant No. FA9550-08-1-0078, FA9550-11-1-0109, FA9550-11-C-0034, FA8650-15-1-7524) and ARPA-E (DE-AR DE-AR0000743) over the past decade.